%% file: main.tex
\documentclass[preprint, 12pt, 3p, numbers, round, sort&compress]{elsarticle}
\makeatletter
\def\ps@pprintTitle{%
 \let\@oddhead\@empty
 \let\@evenhead\@empty
 \def\@oddfoot{}%
 \let\@evenfoot\@oddfoot}
\makeatother

\input{Preamble/dependencies.tex}

\begin{document}

\begin{frontmatter}

\input{Preamble/title_page.tex}

\input{Preamble/Abstract.tex}
\end{frontmatter}

\newpage


\input{Body/body.tex}

\bibliographystyle{ama}
\bibliography{bib.bib}

\pagebreak

\pagebreak

\input{Body/appendix}

\bibliographystyleS{ama}
\bibliographyS{bibAppendix.bib}

\end{document}

%% file: Preamble/dependencies.tex
\usepackage[dvipsnames]{xcolor}
\usepackage{tikz}
\usepackage{longtable,ragged2e,booktabs,siunitx,multirow,lipsum,array, threeparttablex}
\usepackage{natbib}
\usepackage{amssymb}
\usepackage{amsmath}
\usepackage{subcaption}
\usepackage[shortlabels]{enumitem}
\usepackage{mwe}
\usepackage{mathtools}
\usepackage{multicol}
\usepackage{graphicx}
\usepackage{color}
\usepackage[hang, flushmargin, multiple]{footmisc}
\usepackage[hyperfootnotes=false,colorlinks=true]{hyperref}
\usepackage{amsfonts}
\usepackage{graphicx}
\usepackage[utf8]{inputenc}
\usepackage[all]{xy}
\usepackage{csquotes}
\usepackage{ulem}
\usepackage{amsmath}
\usepackage{amsthm}
\usepackage{comment}
\usepackage{bbm}
\usepackage{arydshln}
\usepackage{setspace}
\usepackage[resetlabels]{multibib}
\usepackage{pifont}
\usepackage{soul}
\usepackage{letltxmacro}
\LetLtxMacro{\originaleqref}{\eqref}
\renewcommand*{\eqref}[1]{(equation~\ref{#1})}

\newcommand{\cmark}{\ding{51}}%
\newcommand{\xmark}{\ding{55}}%

\newcolumntype{P}[1]{>{\RaggedRight\arraybackslash\hspace{0pt}}p{#1}}

\newcites{S}{References for Supplemental Materials}

\usetikzlibrary{shapes,decorations,arrows,calc,arrows.meta,fit,positioning}
\tikzset{
    -Latex,auto,node distance =1 cm and 1 cm,semithick,
    state/.style ={ellipse, draw, minimum width = 0.7 cm},
    point/.style = {circle, draw, inner sep=0.04cm,fill,node contents={}},
    bidirected/.style={Latex-Latex,dashed},
    el/.style = {inner sep=2pt, align=left, sloped}
}

%% file: Preamble/title_page.tex
\title{Separable effects for adherence}

\author[1]{Kerollos Nashat Wanis \corref{cor1}}
\author[2]{Mats Julius Stensrud}
\author[2]{Aaron Leor Sarvet}

\address[1]{Division of General Surgery, Western University, Canada}
\address[2]{Department of Mathematics, École Polytechnique Fédérale de Lausanne, Switzerland}
\cortext[cor1]{\textbf{Contact information for corresponding author:}\\
Kerollos N. Wanis, General Surgery, London Health Sciences Centre, Rm. C8-114, London, Ontario, Canada, N6A 5A5. \url{knwanis@gmail.com} \\
Stensrud and Sarvet were funded by the Swiss National Science Foundation, grant $200021\_207436$.  
}

%% file: Preamble/Abstract.tex
\begin{abstract}
\begin{singlespace}
\noindent Comparing different medications is complicated when adherence to these medications differs. We can overcome the adherence issue by assessing effectiveness under sustained use, as in the usual causal `per-protocol' estimand. However, when sustained use is challenging to satisfy in practice, the usefulness of this estimand can be limited. Here we propose a different class of estimands: \textit{separable effects for adherence}. These estimands compare modified medications, holding fixed a component responsible for non-adherence. Under assumptions about treatment components' mechanisms of effect, the separable effects estimand can eliminate differences in adherence. These assumptions are amenable to interrogation by subject-matter experts and can be evaluated using causal graphs. We describe an algorithm for constructing causal graphs for separable effects, illustrate how these graphs can be used to reason about assumptions required for identification, and provide semi-parametric weighted estimators.

\vspace{10pt}

\noindent%

\end{singlespace}
\end{abstract}

%% file: Body/body.tex
\section{Introduction}
\label{section: Introduction}

\noindent Comparing different medications is a core objective in pharmacoepidemiologic studies \citep{Rothman2021Modern}. In these studies, the term efficacy is defined as ``performance of a treatment under ideal and controlled circumstances,'' while effectiveness refers to ``the performance of a
treatment under usual or `real world' circumstances'' \citep{revicki1999pharmacoeconomic}. For sustained pharmacologic treatments, adherence, the continuous utilization of a prescribed medication, is often central to the distinction between efficacy and effectiveness. While a range of adherence strategies can be defined, two types have received special attention: the causal `per-protocol' strategy, which requires continuous adherence, and the `intention-to-treat' strategy, which permits adherence to vary naturally for each individual. 

When continuous adherence is expected for treatments, the usual `per-protocol' strategy is suitable for studying their effectiveness. But in some settings, continuous adherence may be too demanding in practice \citep{osterberg2005adherence, world2003adherence}. In observed data, nonadherence manifests as a lack of support for certain conditional distributions, known as positivity violations \citep{petersen2012diagnosing, kennedy2019nonparametric, robins1986new}. These violations preclude identification of usual `per-protocol' parameters under standard assumptions for causal inference. Further, when non-adherence is prevailing, studying treatments in an idealized setting where everybody adheres has limited relevance to investigators interested in effectiveness.

Conversely, when medication adherence is allowed to vary naturally as in the `intention-to-treat' strategy, a difference in the effectiveness of two initiated medications might arise simply due to a difference in adherence processes \citep{hernan2012beyond, murray2018patients, wanis2022role}. Even when strict utilization strategies are not of primary interest, perhaps because non-adherence is widespread for the treatments under study, investigators often want to compare strategies with equivalent adherence. In other words, in settings with non-adherence, there is interest in a trade-off between the `real-world' adherence that defines effectiveness and the idealized adherence that characterizes efficacy.

These considerations are evident in the comparison of medications for hypertension monotherapy, which we will use as a running example. International and US guidelines recommend initial monotherapy for individuals with low risk grade 1 hypertension to reduce the risk of adverse cardiovascular outcomes \citep{unger20202020, whelton20182017}. Investigators interested in antihypertensive evaluation and refinement will evaluate the comparative effectiveness of available agents. But antihypertensive effectiveness depends on adherence, which is low and varies from agent to agent \citep{ishida2019treatment, elliott2007persistence}.

In this article, we introduce a new estimand for comparative effectiveness research (CER): \textit{separable effects for adherence}. The separable effects for adherence build on a generalized theory of separable effects \citep{robins2022interventionist,robins2010alternative,stensrud2021generalized,stensrud2022separable,stensrud2022conditional}. Their definition requires that investigators conceptualize modifying the medications under study into independently manipulable components, e.g., corresponding to a hypothetical pharmacologic refinement, and consider the components' separate effects on the outcome through adherence pathways, and through other causal pathways. Hypothetical refinements could be considered at the policy level, e.g., a reduction in the cost of the drug, or at the pharmaceutical level, e.g., a change in the size or taste of the drug. In subsequent sections, we elaborate on examples of hypothetical treatment modifications and discuss the identification and interpretation of the separable effect under different assumptions encoded on causal graphs. In some cases, the separable effect quantifies the effectiveness of medication initiation strategies on an outcome of interest under the adherence process of one of the medications.  

In Section~\ref{sec: total and separable effects} we define separable effects, contrasting them with total effects, in Section~\ref{section: causal graphs} we describe the construction of causal graphs for separable effects estimands, in Sections~\ref{section: adherence depends only on Z_A} and \ref{section: multiple_adherence_pathways} we discuss how different assumptions about adherence mechanisms impact the interpretation of the separable effects estimand, in Section~\ref{section: identification} we detail the assumptions required for identification of separable effects using observed data, in Sections~\ref{section: g_formula} and \ref{section: estimation} we discuss algorithms for estimation, and in Section~\ref{section: efficacy} we discuss the role of separable effects estimands in studies of efficacy.

\section{Total and separable effects}
\label{sec: total and separable effects}

\noindent Consider a study where individuals initiate one of two medications. Let $Z$ denote the medication initiated, e.g., let $Z=0$ denote initiation of an angiotensin-converting enzyme inhibitor (ACEI) and $Z=1$ a thiazide diuretic. $Z$ may be randomly assigned, as in an experiment, or selected naturally, as in an observational study. In each interval $k = 1, \ldots, K$, variables are measured: $A_{k}$, an indicator of adherence to the initiated medication; $L_{k}$, covariates (e.g., diagnoses or symptoms); and $Y_{k}$, an indicator of failure (e.g., adverse cardiovascular events). Overlines represent an individual's history through interval $k$ (e.g., $\overline{L}_{k}$). 

Let $Y^{z}_{k}$ denote failure status had, possibly contrary to fact, $Z=z$. The comparative effectiveness of the two medication initiation strategies on the outcome risk by $k$ is defined as

\begin{equation}
    \Pr[Y^{z=1}_{k}=1]\mbox { vs. } \Pr[Y^{z=0}_{k}=1], \label{total effect}
\end{equation} 

\noindent which we refer to as the \textit{total} effect of medication initiation (the `intention-to-treat' effect). 

The total effect of medication initiation depends on adherence, a characteristic that, at least sometimes, is undesirable to investigators, even when comparing `real-world' pharmacologic effectiveness. Suppose that $Z$ can be modified such that it is represented by two binary components, $Z_{A}$ and $Z_{Y}$, i.e., that $Y_{k}^z=Y_{k}^{z_A=z_Y=z}$. With the modified medication, an investigator may consider the effect on an outcome risk of the $Z_{Y}$ component, fixing the $Z_A$ component to the value $z_{A}$,

\begin{equation}
    \Pr[Y^{z_{A}, z_{Y}=1}_{k}=1]\mbox { vs. } \Pr[Y^{z_{A}, z_{Y}=0}_{k}=1], \label{separable effect}
\end{equation} 

\noindent which is the \textit{separable} effect of $Z_{Y}$ initiation under $Z_{A} = z_{A}$. 

An investigator will consider a modification of $Z$ where $Z_A$ is particularly relevant for its effect on adherence. Interventions defining the separable effects correspond to medication modifications that could actually be evaluated in a real-world clinical trial (at least in principle). Arguably, when the $Z_{A}$ component is fixed to a value representing a medication with a more favorable effect on adherence (e.g., a placebo or another well tolerated medication), separable effects estimands can correspond precisely to the parameters that might be observed in experiments conducted for drug development and refinement. As such, they would appeal to investigators interested in emulating this process with data.

Although the choice of an estimand needs to be context specific, separable effects for adherence are useful for pharmacoepidemiologic CER because of the following properties: (a) the resulting adherence patterns are plausible because they arise naturally (that is, investigators do not directly intervene on adherence); (b) their assumptions are amenable to interrogation by subject-matter experts and can be tested (in principle) by experiment; and (c) under assumptions on the mechanisms of $Z_{A}$ and $Z_{Y}$, modified medications will be compared under equivalent adherence. Table~\ref{estimands_table} contrasts separable effects for adherence with other estimands for pharmacoepidemiologic CER.

Property (a) follows immediately because separable effects do not consider interventions under which adherence is directly controlled. In subsequent sections, we will illustrate properties (b) and (c) using causal graphs as tools for reasoning and model representation.

\input{Extra/Tables/estimands_table}

\section{Causal graphs for separable effects}
\label{section: causal graphs}

\noindent Causal inference requires background knowledge. Causal graphs can be constructed based on this knowledge, allowing the use of simple graphical rules to evaluate conditions for identification using data.

Investigators interested in the \textit{total} effect \eqref{total effect} can construct a causal Directed Acyclic Graph (DAG) that includes a single treatment node $Z$ and a single outcome $Y_{k}$. Covariates that are direct causes of any two variables on the graph are then iteratively included, and directed arrows are added between any two variables when a causal effect is supposed to exist. When considering (joint) interventions on more than one variable, an algorithm for constructing the DAG is identical, except initializing the graph with all treatment and outcome nodes, and again iteratively adding all common causes. 

When a separable effects estimand is understood as a total effect of a joint intervention, a DAG can be constructed according to the above classical algorithm. However, even when all common causes of the modified medication components and outcomes are measured, identification by usual strategies will typically fail because the effect involves a combination of components $Z_A$ and $Z_Y$ that have not been implemented in the observed data: whenever $z_A\neq z_Y$, a positivity condition, which is necessary for identification by conventional approaches, will be violated. 

An important contribution of the general literature on separable effects is to clarify graphical conditions for identification via \textit{other} strategies (see, e.g., Robins et al \citep{robins2022interventionist} and Stensrud et al \citep{stensrud2021generalized}). However, these alternate conditions cannot generally be evaluated in graphs constructed by the classical algorithm, because they will involve variables that investigators would not typically be prompted to include on a DAG. However, this classical algorithm is not the \textit{only} approach for constructing valid causal DAGs. A valid DAG may instead be constructed by initializing a graph with not only the treatments and outcome -- here $Z_A$, $Z_Y$, and $Y_{k}$ -- but also with a set of intervening (mediating) variables that, minimally, are involved in the causal mechanism of the effect of $Z_A$ on $Y_{k}$. With separable effects for adherence, this set will include the adherence indicators $\overline{A}_k$. Furthermore, this initial set of intervening variables should be expanded if there still remain direct paths from $Z_A$ into the outcome $Y_{k}$ or direct paths from $Z_Y$ into any adherence indicator in $\overline{A}_k$. Ultimately, identification via the strategies in \citep{stensrud2021generalized} requires independencies encoded by a causal DAG where the effects of the modified medication components are completely intersected by non-overlapping (separate) sets of intervening variables, except for $Z_Y$, which may have a direct effect on $Y_{k}$. 

In Sections \ref{section: adherence depends only on Z_A} and \ref{section: multiple_adherence_pathways}, we review several classes of graphs for separable effects and discuss their implications for identification and interpretation. To avoid clutter, we restrict the graphs to two time points and assume the initiated medication, $Z$, was randomly assigned, as in a clinical trial. Likewise, we suppose that no outcome events occur in the first interval, so that $Y_{1}$ is omitted from the graphs.

\section{Structures where separable effects balance adherence}
\label{section: adherence depends only on Z_A}

\noindent Suppose that the $Z_{Y}$ component of an individual's initiated medication exerts no effect on adherence, neither directly, nor through any covariates, except for past survival. This is a condition that balances adherence for survivors under the separable effect of $Z_{Y}$ under $Z_{A} = z_{A}$. The condition holds in the causal graphs of Figures~\ref{fig: cost_example_graphs}, \ref{fig: nonprognostic_side_effects_example_graphs}, and \ref{fig: prognostic_side_effects_example_graphs} which encode different assumptions about the mechanisms through which the outcome is affected by the modified medication. 

Figure~\ref{fig: cost_example_graphs} represents a setting where differences in adherence are entirely due to the modified medication component $Z_{A}$, which exerts its effect on the outcome only through adherence. This setting can also be represented by Figure~\ref{fig: nonprognostic_side_effects_example_graphs}, which introduces $V_{k}$, a vector of \textit{non-prognostic covariates}. The term \textit{non-prognostic} is used to denote covariates not associated with the outcome except through the initiated medication or adherence. When the $Z_{Y}$ component exerts no effect on adherence, except via past survival, adjustment for non-prognostic variables is not required for identification, but their inclusion on graphs is helpful for reasoning about the underlying assumptions.

Building on our example, suppose $Z$ indicates initiation of an ACEI versus a thiazide diuretic, and investigators consider a modified medication where $Z_A$ represents the medication's out-of-pocket cost \citep{hassan2006identification}, and $Z_Y$ represents the remaining features of the medication. Thiazide diuretics are less costly than ACEIs, and these costs are known to affect adherence in healthcare settings without universal coverage \citep{fischer2004economic, fretheim2003potential, park2020uses, johansen2021total, khera2019cost}. This example is represented by Figure~\ref{fig: cost_example_graphs} if out-of-pocket costs were entirely responsible for adherence differences (alternatively, by Figure~\ref{fig: nonprognostic_side_effects_example_graphs} with current wealth included in the covariate vector $V_{k}$). For this modified medication example, an experiment comparing the effectiveness of the $Z_{Y}$ components of a thiazide diuretic versus an ACEI while setting $Z_{A} = 1$ could be conducted if the out-of-pocket costs of ACEIs were lowered to that of thiazides through pharmaceutical innovation or health policy interventions. 

Alternatively, consider a setting with no effect of differential drug costs on adherence, which, for example, is plausible in a clinical trial where investigators cover drug costs. Suppose the investigators consider a $V_{k}$ representing a history of cough, headache, or angioedema, which are considered to be non-prognostic adverse-effects of ACEIs \citep{gregoire2001tolerability}, but not generally of thiazides. Instead, thiazides can cause other non-prognostic adverse-effects, which are also included in $V_{k}$: urinary frequency, erectile dysfunction, fatigue, and muscle cramps \citep{kronish2011meta}. A version of an initiated medication -- modified to affect non-prognostic adverse-effects in the same way as the other medication -- can be represented by Figure~\ref{fig: nonprognostic_side_effects_example_graphs}. For example, if instead of choosing initiation of a thiazide as the referent drug, investigators chose a drug with minimal adverse-effects, they would arguably be emulating the type of innovation that has already been attempted in the form of angiotensin II receptor blockers, which were developed to overcome some deficiencies in ACEIs \citep{barreras2003angiotensin}. 

\input{Extra/Figures/cost_example_graphs.tex}

\input{Extra/Figures/nonprognostic_side_effects_example_graphs.tex}

So far, we have only considered settings where non-prognostic covariates affect adherence. However, some adverse-effects of antihypertensives are prognostic. ACEIs are thought to cause acute kidney injury (AKI), while the relationship between thiazides and AKI is less clear \citep{albasri2021association, ejaz2014diuretics, nigwekar2011diuretics}. Individuals who develop an AKI are likely to discontinue their initiated medication. Unlike non-prognostic covariates, AKI may increase the risk of cardiovascular disease \citep{legrand2020cardiovascular}. Now, suppose that Figure~\ref{fig: prognostic_side_effects_example_graphs} represents another separable effect setting, with $L_{k}$ representing a history of AKI. Consider an investigator who wants to study the outcome risk under a modification of ACEIs that leads to a similar distribution of AKI as thiazides. In this context, the effect of $Z_A$ on the outcome is not mediated entirely by adherence: there is an effect mediated by $L_2$, as shown in Figure~\ref{fig: prognostic_side_effects_example_graphs}. Nevertheless, the separable effect will contrast treatments under an equivalent distribution of adherence.

\input{Extra/Figures/prognostic_side_effects_example_graphs.tex}

\section{Structures where separable effects do not balance adherence}
\label{section: multiple_adherence_pathways}

\noindent Now consider a setting where the investigators believe that the $Z_{Y}$ component of the modified medication affects adherence. For example, investigators considering a modification of ACEIs equalizing their out-of-pocket costs relative to thiazides will have this belief if prognostic covariates, such as AKI, affect adherence, as represented in Figure~\ref{fig: prognostic_side_effects_example_graphs2}. 

Suppose these same investigators study a modified medication that not only equalizes out-of-pocket costs, but also the distribution of AKI. This setting can be represented by Figure~\ref{fig: prognostic_side_effects_example_graphs}. However, even in this setting, the investigators may not accept that adherence is unaffected by $Z_{Y}$ if there are other prognostic covariates that could affect adherence. In some cases, the effect of $Z_{Y}$ on the outcome will be mediated by measured intermediaries whose values will be known to the individuals under treatment and might therefore impact adherence (e.g., measured blood pressure \citep{rahimi2021pharmacological}). In these cases, both components of a modified medication $Z_{A}$ and $Z_{Y}$ will exert effects on adherence, through different sets of prognostic covariates. The investigators can represent this setting using Figure~\ref{fig: subset_prognostic_covariates}. The prognostic covariates $L_{k}$ have been divided into two subvectors $L_{A,k}$ (which includes AKI) and $L_{Y,k}$ (which includes systolic and diastolic blood pressure), and both $Z_{Y}$ and $Z_{A}$ exert effects on the outcome through their effects on adherence.  

In both Figures~\ref{fig: prognostic_side_effects_example_graphs2} and \ref{fig: subset_prognostic_covariates} the separable effect \eqref{separable effect} considers a contrast that eliminates some, but not all, differences in adherence. However, the extent to which adherence to antihypertensives depends on blood pressure rather than the convenience, cost, and tolerability of the initiated medication is open to debate by subject-matter experts \citep{burnier2019adherence}. Whether our investigators are correct about the impact of blood pressure on adherence is important for deciding whether their modified medication that balances the distribution of cost and AKI is representable by Figure~\ref{fig: prognostic_side_effects_example_graphs} or must be represented by Figure~\ref{fig: subset_prognostic_covariates}, and thus whether their estimand balances adherence. 

\input{Extra/Figures/prognostic_side_effects_example_graphs2.tex}

\input{Extra/Figures/subset_prognostic_covariates.tex} 

\section{Identification of separable effects using observed data}
\label{section: identification}

\noindent Graphical conditions for identification of $\Pr[Y^{z_{A},z_{Y}}_{K}=1]$ differ from those classically used for identification of $\Pr[Y^{z}_{K}=1]$. In the identification strategies we review for each of these estimands, we require unconfoundedness for the initiated medication $Z$, which can be read from a DAG as the absence of backdoor paths connecting $Z$ and $Y_{K}$. This condition is expected to hold in data from a study where $Z$ is randomly assigned, or may hold conditional on baseline covariates in an observational study.

Identification of $\Pr[Y^{z_{A},z_{Y}}_{K}=1]$ further requires a set of conditions not required for identification of $\Pr[Y^{z}_{K}=1]$ \citep{stensrud2021generalized}. Specifically, we prohibit any directed paths connecting $Z_{A}$ and $Y_{k}$ or $L_{Y,k}$ conditional on the intervening treatment, covariate, and outcome history from $k$, for each $k = 1, \ldots, K$. Analogously, we prohibit directed paths \citep{greenland1999causal} connecting $Z_{Y}$ and $A_{k}$ or $L_{A,k}$ conditional on the intervening treatment, covariate, and outcome history from $k$, for each $k = 1, \ldots, K$. It is possible that these conditions are violated even when the data arise from a study where $Z$ is randomly assigned. Section~\ref{section: exchangeability violated} provides examples where these assumptions would not hold.

Investigators should be mindful that features of the causal structure represented by the graph for the observed data can change in the contexts considered by separable effects estimands, that is, when $z_A\neq z_Y$ \citep{robins2022interventionist,stensrud2022conditional}. For example, a variable that is not a confounder in the observed data setting, because it is a cause of the initiated medication, $Z$, but not an outcome $Y_{k}$, might cause the outcome in a setting where $z_A \neq z_Y$, violating unconfoundedness for the initiated medication. Similarly, a variable that was solely a cause of adherence in the observed data setting may also be a cause of the outcome under $z_A \neq z_Y$, violating an independence condition not required for classical identification strategies but necessary for separable effects. These theoretical issues will not arise if the causal structure between variables in the graph is correctly assumed to be invariant from the observed data context to the separable effect contexts considered.

\subsection{When do independence conditions for separable effects not hold?}
\label{section: exchangeability violated}

\noindent To make transparent the assumptions required for identification of the separable effect, we now give three examples of their violation. 

First, suppose, as we did in Figure~\ref{fig: prognostic_side_effects_example_graphs}, that $\overline{L}_{k}$ includes a history of prognostic variables that have an effect on adherence. But also suppose, unlike in Figure~\ref{fig: prognostic_side_effects_example_graphs}, that there are unmeasured common causes of prognostic variables and outcomes. This is depicted in Figure~\ref{fig: prognostic_side_effects_example_graphs_not_identified}a. The path $Z_{A} \rightarrow \boxed{L_{A,2}} \leftarrow U \rightarrow Y_{2}$, which is unblocked due to conditioning (denoted by enclosing a variable in a square) on the collider $L_{A,2}$ \citep{greenland1999causal}, violates an independence assumption required for identification. Similarly, identification will fail whenever there is an unmeasured common cause of $L_{Y,k}$ and any future treatment $A_{k'}$, $k'>k$. More generally, identification will fail if there are any unmeasured common causes of an element in $\{\overline{A}_K, \overline{L}_{A,K}\}$ and an element in $\{\overline{Y}_K, \overline{L}_{Y,K}\}$.

Independence assumptions can still be violated even in the absence of unmeasured common causes. Figure~\ref{fig: prognostic_side_effects_example_graphs_not_identified}b illustrates a violation due to an unmeasured intermediate on the causal path from $Z_{A}$ to $Y_{2}$ and Figure~\ref{fig: prognostic_side_effects_example_graphs_not_identified}c shows an analogous violation due to an unmeasured intermediate on the causal path from $Z_{Y}$ to $L_{A,2}$. These latter two examples illustrate the importance of collecting information on potential mediating variables whenever separable effects estimands are targeted.  More generally, identification will fail if there is any unblocked path from $Z_Y$ into $\{\overline{A}_K, \overline{L}_{A,K}\}$ or if there is any unblocked path from $Z_A$ into $\{\overline{Y}_K, \overline{L}_{Y,K}\}$. 
 
\input{Extra/Figures/prognostic_side_effects_example_graphs_not_identified.tex}

\section{The g-formula}
\label{section: g_formula}

\noindent Suppose that the investigator articulates a realistic modified medication with components $Z_{A}$ and $Z_{Y}$, and that consistency, which allows linking counterfactual to factual random variables, holds. Suppose also that the distribution of variables on the causal DAG is positive. When suitable independence conditions hold -- implied by the absence of biasing paths in a DAG as discussed in Section~\ref{section: identification} -- then the expected counterfactual mean of $Y_{K}$ under an intervention that sets $Z_{Y} = z_{Y}$ and $Z_{A} = z_{A}$ is given by the g-formula \citep{stensrud2021generalized},

\begin{align}
    \sum_{k=1}^{K}\sum_{\overline{a}_k}
    &\sum_{\overline{l}_{k}} \Pr[Y_{k} = 1 \mid \overline{A}_{k} = \overline{a}_{k}, \overline{L}_{Y,k} = \overline{l}_{Y,k}, \overline{L}_{A,k} = \overline{l}_{A,k}, \overline{Y}_{k-1}=0, Z=z_{Y}] \times \nonumber \\
    \prod_{j=1}^{k} \{&\Pr[Y_{j-1} = 0 \mid \overline{A}_{j-1} = \overline{a}_{j-1}, \overline{L}_{Y,j-1} = \overline{l}_{Y,j-1}, \overline{L}_{A,j-1} = \overline{l}_{A,j-1}, \overline{Y}_{j-2}=0, Z=z_{Y}] \times \label{gformula} \\
    &f(l_{Y,j} \mid \overline{a}_{j-1}, \overline{l}_{Y,j-1}, \overline{l}_{A,j}, \overline{Y}_{j-1} = 0, Z=z_{Y}) \times \nonumber \\
    &f(l_{A,j} \mid \overline{a}_{j-1}, \overline{l}_{Y,j-1}, \overline{l}_{A,j-1}, \overline{Y}_{j-1} = 0, Z=z_{A}) \times \nonumber \\
    &f(a_j \mid  \overline{a}_{j-1}, \overline{l}_{Y,j}, \overline{l}_{A,j}, \overline{Y}_{j-1} = 0, Z=z_{A})\} \nonumber.
\end{align}

This g-formula can be re-expressed as a weighted representation,

\begin{align*}
    \sum_{k=1}^{K} \lambda_{A,k} \prod_{j=1}^{k-1} [1-\lambda_{A,j}] \text{ or } \sum_{k=1}^{K} \lambda_{Y,k} \prod_{j=1}^{k-1} [1-\lambda_{Y,j}]
\end{align*}

where

\begin{align*}
    \lambda_{A,k} = \frac{\mathrm{E}[Y_k(1-Y_{k-1})W_{A,k} \mid Z=z_{Y}]}{\mathrm{E}[(1-Y_{k-1})W_{A,k} \mid Z=z_{Y}]} \text{ and } \lambda_{Y,k} = \frac{\mathrm{E}[Y_k(1-Y_{k-1})W_{Y,k} \mid Z=z_{A}]}{\mathrm{E}[(1-Y_{k-1})W_{Y,k} \mid Z=z_{A}]} 
\end{align*}

and the weights $W_{A,k}$ and $W_{Y,k}$ are defined as

\begin{align*}
W_{A,k}= \prod_{j=1}^{k} \Bigg[ &\frac{f(A_{j} \mid \overline{Y}_{j-1}=0,\overline{L}_{Y,j},\overline{L}_{A,j},\overline{A}_{j-1}, Z=z_{A})}{f(A_{j} \mid \overline{Y}_{j-1}=0,\overline{L}_{Y,j},\overline{L}_{A,j},\overline{A}_{j-1}, Z=z_{Y})} \times \\
& \frac{\Pr(Z=z_{A} \mid \overline{Y}_{j-1}=0,\overline{L}_{Y,j-1},\overline{L}_{A,j},\overline{A}_{j-1})}{\Pr(Z=z_{Y} \mid \overline{Y}_{j-1}=0,\overline{L}_{Y,j-1},\overline{L}_{A,j},\overline{A}_{j-1})} \times \\
& \frac{\Pr(Z=z_{Y} \mid \overline{Y}_{j-1}=0,\overline{L}_{Y,j-1},\overline{L}_{A,j-1},\overline{A}_{j-1})}{\Pr(Z=z_{A} \mid \overline{Y}_{j-1}=0,\overline{L}_{Y,j-1},\overline{L}_{A,j-1},\overline{A}_{j-1})}\Bigg]
\end{align*}

and

\begin{align*}
W_{Y,k}= & \prod_{j=1}^{k}\frac{f(Y_{j} \mid \overline{Y}_{j-1}=0, \overline{L}_{Y,j},\overline{L}_{A,j},\overline{A}_{j}, Z=z_{Y})}{f(Y_{j} \mid \overline{Y}_{j-1}=0, \overline{L}_{Y,j},\overline{L}_{A,j},\overline{A}_{j},Z=z_{A})} \times \\
& \prod_{j=1}^{k} \Bigg[\frac{\Pr(Z=z_{Y} \mid \overline{Y}_{j-1}=0,\overline{L}_{Y,j},\overline{L}_{A,j},\overline{A}_{j-1})}{\Pr(Z=z_{A} \mid \overline{Y}_{j-1}=0,\overline{L}_{Y,j},\overline{L}_{A,j},\overline{A}_{j-1})} \times \\
& \hspace{1cm} \frac{\Pr(Z=z_{A} \mid \overline{Y}_{j-1}=0,\overline{L}_{Y,j-1},\overline{L}_{A,j},\overline{A}_{j-1})}{\Pr(Z=z_{Y} \mid \overline{Y}_{j-1}=0,\overline{L}_{Y,j-1},\overline{L}_{A,j},\overline{A}_{j-1})}\Bigg].
\end{align*}

The equivalence of these expressions is shown in Supplementary Materials Sections~\ref{section: IPW proof} and \ref{section: alternative IPW proof}. Weighted representations motivate inverse probability weighted (IPW) estimators. Choosing between the two weighted representations should not be done arbitrarily, and should be guided by assumptions about the data generating mechanism producing the study data, which can be represented on causal graphs. When $L_{k}=(L_{Y,k}, \emptyset), k=1,\ldots,K$, then $W_{A,k}$ simplifies to 

\begin{align*}
\prod_{j=1}^{k} \frac{f(A_{j} \mid \overline{Y}_{j-1}=0,\overline{L}_{Y,j},\overline{L}_{A,j},\overline{A}_{j-1}, Z=z_{A})}{f(A_{j} \mid \overline{Y}_{j-1}=0,\overline{L}_{Y,j},\overline{L}_{A,j},\overline{A}_{j-1}, Z=z_{Y})}
\end{align*}

\noindent because there are no variables in $L_{A,k}$ needed for identification (e.g., Figure~\ref{fig: prognostic_side_effects_example_graphs2}).

Similarly, when $L_{k}=(\emptyset, L_{A,k}), k=1,\ldots,K$ then $W_{Y,k}$ simplifies to

\begin{align*}
\prod_{j=1}^{k} \frac{f(Y_{j} \mid \overline{Y}_{j-1}=0, \overline{L}_{Y,j},\overline{L}_{A,j},\overline{A}_{j}, Z=z_{Y})}{f(Y_{j} \mid \overline{Y}_{j-1}=0, \overline{L}_{Y,j},\overline{L}_{A,j},\overline{A}_{j},Z=z_{A})}
\end{align*}

\noindent because, likewise, there are no variables in $L_{Y,k}$ needed for identification (e.g., Figure~\ref{fig: prognostic_side_effects_example_graphs}).

The weighted representation that includes the most simplified expression will be preferred, because estimation based on this representation will require the fewest correctly specified models. When neither expression can be simplified, the choice of the representation should be justified using substantive reasoning about which models can be specified correctly. For example, investigators may feel more confident specifying parametric models for adherence rather than for outcomes. 

For technical articulations of analagous identification conditions, see \citep{stensrud2021generalized}.

\section{Inverse probability weighted estimation}
\label{section: estimation}

\noindent If the terms in $W_{Y,k}$, can be modelled correctly using parametric methods, then the following IPW algorithm will be a consistent estimator for $\Pr[Y_K^{z_A, z_Y}=1]$, with $z_A \neq z_Y$:

\begin{footnotesize}

\begin{enumerate} 
    \item Using the entire data set, fit a pooled (over time) parametric regression model for 
    \begin{itemize}
    \item $\Pr(Y_{k} = y_{k} \mid \overline{Y}_{k-1}=0, \overline{l}_{Y,k},\overline{l}_{A,k},\overline{a}_{k}, z)$,
    \item $\Pr(Z = z \mid \overline{Y}_{k-1}=0,\overline{l}_{Y,k},\overline{l}_{A,k},\overline{a}_{k-1})$, and
    \item $\Pr(Z = z \mid \overline{Y}_{k-1}=0,\overline{l}_{Y,k-1},\overline{l}_{A,k},\overline{a}_{k-1})$.
    \end{itemize}
    For example, we might assume pooled logistic regression models. 
    \item For each row in the data set for each individual with $Z=z_{A}$, at each time interval, $k$, in $1, \ldots, K$:
    \begin{enumerate}
        \item Obtain predicted values 
        \begin{itemize}
            \item $\widehat{f}(Y_{k} \mid \overline{Y}_{k-1}=0, \overline{L}_{Y,k},\overline{L}_{A,k},\overline{A}_{k}, Z=z_{A})$,
            \item $\widehat{f}(Y_{k} \mid \overline{Y}_{k-1}=0, \overline{L}_{Y,k},\overline{L}_{A,k},\overline{A}_{k}, Z=z_{Y})$,
            \item $\widehat{\Pr}(Z = z_{A} \mid \overline{Y}_{k-1}=0,\overline{L}_{Y,k-1},\overline{L}_{A,k},\overline{A}_{k-1})$,
            \item $\widehat{\Pr}(Z = z_{Y} \mid \overline{Y}_{k-1}=0,\overline{L}_{Y,k-1},\overline{L}_{A,k},\overline{A}_{k-1})$,
            \item $\widehat{\Pr}(Z = z_{A} \mid \overline{Y}_{k-1}=0,\overline{L}_{Y,k-1},\overline{L}_{A,k-1},\overline{A}_{k-1})$, and
            \item $\widehat{\Pr}(Z = z_{Y} \mid \overline{Y}_{k-1}=0,\overline{L}_{Y,k-1},\overline{L}_{A,k-1},\overline{A}_{k-1})$.
        \end{itemize}
        
        \item Evaluate 
        
        \begin{enumerate}
         \item 
         \begin{align*}
            \widehat{W}_{Y,k}= &\prod_{j=1}^{k}\frac{\widehat{f}(Y_{j} \mid \overline{Y}_{j-1}=0, \overline{L}_{Y,j},\overline{L}_{A,j},\overline{A}_{j}, Z=z_{Y})}{\widehat{f}(Y_{j} \mid \overline{Y}_{j-1}=0, \overline{L}_{Y,j},\overline{L}_{A,j},\overline{A}_{j},Z=z_{A})} \times \\
            & \prod_{j=1}^{k} \Bigg[\frac{\widehat{\Pr}(Z=z_{Y} \mid \overline{Y}_{j-1}=0,\overline{L}_{Y,j},\overline{L}_{A,j},\overline{A}_{j-1})}{\widehat{\Pr}(Z=z_{A} \mid \overline{Y}_{j-1}=0,\overline{L}_{Y,j},\overline{L}_{A,j},\overline{A}_{j-1})} \times \\
            & \hspace{1cm} \frac{\widehat{\Pr}(Z=z_{A} \mid \overline{Y}_{j-1}=0,\overline{L}_{Y,j-1},\overline{L}_{A,j},\overline{A}_{j-1})}{\widehat{\Pr}(Z=z_{Y} \mid \overline{Y}_{j-1}=0,\overline{L}_{Y,j-1},\overline{L}_{A,j},\overline{A}_{j-1})}\Bigg]
        \end{align*}
        or, when $L_{k}=(\emptyset, L_{A,k}), k=1,\ldots,K$, (e.g., Figure~\ref{fig: prognostic_side_effects_example_graphs2}), the simplified weight expression
        \item 
        \begin{align*}
            \widehat{W}_{Y,k}= &\prod_{j=1}^{k}\frac{\widehat{f}(Y_{j} \mid \overline{Y}_{j-1}=0, \overline{L}_{Y,j},\overline{L}_{A,j},\overline{A}_{j}, Z=z_{Y})}{\widehat{f}(Y_{j} \mid \overline{Y}_{j-1}=0, \overline{L}_{Y,j},\overline{L}_{A,j},\overline{A}_{j},Z=z_{A})}.
        \end{align*}
        \end{enumerate}
    \end{enumerate}
    \item Compute the risk of failure by the end of $K$ as $\sum_{k=1}^{K} \widehat{\lambda}_{Y,k} \displaystyle  \prod_{j=1}^{k-1} [1-\widehat{\lambda}_{Y,j}]$ with the estimated weights, $\widehat{W}_{Y,k}$, in $\widehat{\lambda}_{Y,k} = \frac{\widehat{E}[Y_k(1-Y_{k-1})\widehat{W}_{Y,k} \mid Z=z_{A}]}{\widehat{E}[(1-Y_{k-1})\widehat{W}_{Y,k} \mid Z=z_{A}]}$.
\end{enumerate}

\end{footnotesize}

In a study of two medications, to compute the separable effect of $Z_{Y}$ setting $Z_{A}=z_{A}$, the analyst would compute an estimate of the outcome under $z_{Y}\neq z_{A}$ using the IPW algorithm and compare it to the `intention-to-treat' estimate of the outcome under $Z=z_{A}$, which can be computed as a simple empirical mean of $Y_K$ among individuals with $Z=z_{A}$.

An analogous algorithm using $W_{A,k}$ is detailed in Supplementary Materials Section~\ref{section: alternative_IPW}. These algorithms assume $Z$ was randomly assigned. When data arises from an observational study, further adjustment for baseline confounding will be necessary. 

Valid 95\% confidence intervals can be obtained using a non-parametric bootstrap.

In Supplementary Materials Section~\ref{section: toy_example} we implement the algorithm in a simulated data example.

\section{Separable effects and `efficacy'}
\label{section: efficacy}

\noindent In Section~\ref{section: identification}, we discussed assumptions under which the g-formula \eqref{gformula} identifies terms of the separable effect \eqref{separable effect} of $Z_{Y}$ under $Z_{A}=z_{A}$. An advantage of this estimand, in contrast to the total effect \eqref{total effect} of $Z$, is that its value cannot be due to differential adherence to the initiated medications whenever certain conditions hold, which can be encoded on causal graphs (Section~\ref{section: adherence depends only on Z_A}). This is useful when investigators are interested in medication effectiveness comparisons that preclude certain adherence-based mechanisms, which can intuitively be interpreted as a trade-off between the usual `intention to treat' effectiveness comparison versus an efficacy comparison specifying perfect adherence. 

Efficacy refers to ``biomedical end-points ... under optimal
and highly controlled experimental conditions'' and is achieved by studying medications ``under highly unusual and structured protocols by very motivated
clinicians with careful monitoring'' \citep{revicki1999pharmacoeconomic}. Despite balancing adherence, separable effects  will generally not be well suited to studying efficacy, because they will not satisfy all optimal conditions, even those related to adherence-based \textit{mechanisms}. Specifically, consider a setting where all individuals who do not adhere under $Z_Y=1, Z_A=z_A$ are simply untreated for their hypertension during those time-points, whereas all individuals who do not adhere under $Z_Y=0, Z_A=z_A$ instead switch (or `cross-over') to the pharmacologic treatment consistent with $Z_Y=1$. Clearly, even if adherence patterns are nominally balanced (according to the definition of the vector $\overline{A}_K$), an investigator who is purely interested in the \textit{efficacy} of the $Z_Y$ component will be less interested in the separable effect of $Z_{Y}$ under $Z_{A}=z_{A}$ when such data generating mechanisms are possible.      

To preclude such data generating mechanisms, we could restrict ourselves to studies where individuals are unable to take any of the study medications except the one they were initially assigned. A randomized trial where participants only have access to their assigned medication, and not to any of the other study drugs, is an example of such a study. But, in many cases, investigators will be interested in analyzing data arising from an observational study, or from a `pragmatic' randomized trial, where individuals have access to medications other than the one they initiated at baseline. In such settings, a separable effects estimand under a simultaneous hypothetical intervention to prohibit `crossing-over' can be considered. This alternative estimand is discussed in Supplementary Materials Section~\ref{section: switching prohibited}.

Even when crossovers to non-initiated medications are prohibited, separable effects estimands may still not satisfy investigators seeking to emulate the optimal conditions required for studying efficacy. Investigators will need to be precise about the particular mediating pathways, where adherence and crossover are just two of potentially many others, that ought to be excluded in order for the effect estimate to have the desired efficacy interpretation. For example, suppose that the investigators define the relative efficacy of two medications as a contrast of protocols that not only prohibit co-medications and rescue treatments, but also specify the lifestyle choices allowable under the treatment strategies. These various specifications imply that efficacy can refer to a variety of estimands that prohibit different types of mechanisms, depending on the particular interpretation desired.

\section{Discussion}

\noindent We described an estimand, based on the generalized theory of separable effects, that can compare medication initiation strategies while eliminating differences in adherence under assumptions that are well suited to interrogation by subject-matter experts.  

Investigators who are naive to separable effects estimands might have considered different estimands based on the property of balanced adherence. Natural direct effects would balance adherence by assigning an adherence pattern to each individual that is precisely equal to the pattern that would arise for that individual under $Z=z_A$ \citep{vanderweele2017mediation}. Randomized interventional analogues assign an adherence pattern from the distribution of patterns under $Z=z_A$ \citep{vanderweele2017mediation}. Controlled direct effects can balance adherence via interventions, either through enforcement of continuous treatment adherence (as in `per-protocol' estimands) or by imposing an investigator-selected distribution of adherence \citep{wanis2022role} via a stochastic intervention. Lastly, principal stratum direct effects can balance adherence by considering the subgroup of the population who would have (counterfactually) continuously adhered regardless of initiated medication \citep{bornkamp2021principal}. Each of these other estimands have some but not all the properties of separable effects (see Table~\ref{estimands_table}). In contrast to these other estimands, the separable effects contrast does not guarantee equivalent adherence by its definition. Instead, equivalent adherence depends on assumptions about the treatment components' mechanisms of effect, which may be encoded in causal DAGs. Ultimately, choosing the appropriate estimand will be based on deep interdisciplinary dialogue, with subject matter expertise, specific to the particular treatments under comparison.

We used an example of antihypertensive medications to illustrate important assumptions for identification of separable effects, but these assumptions may hold in many other settings. For example, antiretroviral drugs \citep{tseng2020medicare}, antiepileptics \citep{callaghan2019out}, and antihyperglycemics \citep{bibeau2016impact, dejong2020out} are each sustained use medication classes with substantial variability in out-of-pocket costs. A comparison of medications within one of these classes using a separable effects estimand could be consistent with the assumptions discussed in Section~\ref{section: adherence depends only on Z_A} if differences in adherence are driven by differences in costs. Further, DAGs similar to Figures~\ref{fig: cost_example_graphs} and \ref{fig: nonprognostic_side_effects_example_graphs} could arise in settings where investigators consider a modified medication, manipulating a $Z_A$ denoting the complexity of the dosing, the taste or size of the drug, the logistics of its dispensing, or the component that exerts effects on non-prognostic covariates. Investigators can reason about whether assumptions allowing identification of the separable effect hold by constructing causal graphs, guided by pharmacoepidemiologic expertise.

%% file: Extra/Tables/estimands_table.tex
\begin{table}[htbp]
\caption{Properties of selected estimands that compare the effectiveness of pharmacologic treatments.}
\begin{centering}
\begin{footnotesize}
\begin{tabular}{
>{\raggedright\arraybackslash}p{3.3cm}
>{\raggedright\arraybackslash}p{6.7cm}
>{\centering\arraybackslash}p{1.53cm}
>{\centering\arraybackslash}p{1.2cm}
>{\centering\arraybackslash}p{1.53cm}
}
\toprule
\toprule
Estimand & Description & Natural adherence$^{a}$ & Testable$^{b}$ & Equivalent adherence$^{c}$ \\
\midrule
\midrule
`Intention-to-treat' effect \citep{gupta2011intention} & Comparison of risks under the natural adherence processes of the medications & \textcolor{OliveGreen}{\cmark} & \textcolor{OliveGreen}{\cmark} & \textcolor{BrickRed}{\xmark} \\\\
`Per-protocol' effect \citep{hernan2012beyond} & Comparison of risks under interventions that enforce continuous adherence & \textcolor{BrickRed}{\xmark} & \textcolor{OliveGreen}{\cmark} & \textcolor{OliveGreen}{\cmark} \\\\
Principal stratum direct effect \citep{bornkamp2021principal} & Comparison of risks in the subgroup of the population who would continuously adhere under either medication & \textcolor{OliveGreen}{\cmark} & \textcolor{BrickRed}{\xmark} & \textcolor{OliveGreen}{\cmark} \\\\
Natural direct effect$^{d}$ \citep{vanderweele2017mediation} & Comparison of risks under interventions that set adherence for each individual to its counterfactual value under one of the medications & \textcolor{BrickRed}{\xmark} & \textcolor{BrickRed}{\xmark} & \textcolor{OliveGreen}{\cmark} \\\\
Randomized interventional analogue of the natural direct effect \citep{vanderweele2017mediation} & Comparison of risks under the counterfactual distribution of adherence that would arise under one of the medications & \textcolor{BrickRed}{\xmark} & \textcolor{OliveGreen}{\cmark} & \textcolor{OliveGreen}{\cmark} \\\\
Stochastic adherence effect \cite{wanis2022role} & Comparison of risks under interventions that randomly determine adherence according to an investigator-specified distribution & \textcolor{BrickRed}{\xmark} & \textcolor{OliveGreen}{\cmark} & \textcolor{OliveGreen}{\cmark} \\\\
Separable effect & Comparison of risks under the natural adherence processes of modified medications, holding fixed a component responsible for non-adherence & \textcolor{OliveGreen}{\cmark} & \textcolor{OliveGreen}{\cmark} & \textcolor{OliveGreen}{\cmark}$^{e}$ \\
\bottomrule
\bottomrule
\end{tabular}
\label{estimands_table} 
\end{footnotesize}
\end{centering}

\noindent \footnotesize{$a$ Investigators do not directly intervene on adherence. \\
$b$ Assumptions can be tested (in principle) by experiment. \\
$c$ Medications are compared under equivalent adherence. \\
$d$ Natural direct effects are not identified when there are time-varying confounders affected by prior exposure and mediator \cite{vanderweele2017mediation}. \\
$e$ The equivalent adherence expected under the interventions comprising the separable effects contrast does not follow by definition of the parameter but is rather contingent on assumptions, which may be encoded in causal DAGs, as in Section~\ref{section: adherence depends only on Z_A}. \\}
\end{table}

%% file: Extra/Figures/cost_example_graphs.tex
\begin{figure}
    \centering
\scalebox{0.65}{
\begin{tikzpicture}
\begin{scope}[every node/.style={thick,draw=none}]

    \node (Z)   at ( 0, 0 ) {$Z$};
    \node (ZY)   at ( 1.5, 1 ) {$Z_Y$};
    \node (ZA)   at ( 1.5, -1 ) {$Z_A$};

    \node (A1)   at ( 4, 0 ) {$A_1$};
    \node (A2)  at ( 7, 0 ){$A_2$};

    \node (Y)   at ( 10, 0 ) {$Y_{2}$};
\end{scope}
\begin{scope}[>={Stealth[black]},
              every node/.style={fill=white,circle},
              every edge/.style={draw=black,very thick}]

    \path [->] (Z)  edge (ZA);
    \path [->] (Z)  edge (ZY);

    \path [->] (ZY)  edge[bend left=15] (Y);

    \path [->] (ZA)  edge (A1);
    \path [->] (ZA)  edge[bend right=15] (A2);
        
    \path [->] (A1)  edge (A2);
    
    \path [->] (A1)[bend right=25]  edge (Y);
    \path [->] (A2)  edge (Y);
\end{scope}
\end{tikzpicture}
}
\caption{Causal DAG representing a setting in which the $Z_{A}$ component of a modified treatment exerts an effect on $Y_{2}$ only through adherence, and the $Z_{Y}$ component exerts no effect on adherence.}
\label{fig: cost_example_graphs}
\end{figure}
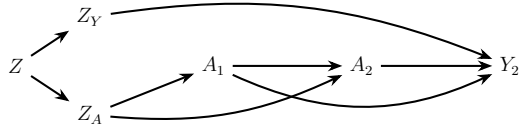

%% file: Extra/Figures/nonprognostic_side_effects_example_graphs.tex
\begin{figure}
    \centering
\scalebox{0.65}{
\begin{tikzpicture}
\begin{scope}[every node/.style={thick,draw=none}]

    \node (V1)   at ( 3, -3 ) {$V_1$};
    \node (V2)  at ( 6, -3 ) {$V_2$};

    \node (Z)   at ( 0, 0 ) {$Z$};
    \node (ZY)   at ( 1.5, 1 ) {$Z_Y$};
    \node (ZA)   at ( 1.5, -1 ) {$Z_A$};

    \node (A1)   at ( 4, 0 ) {$A_1$};
    \node (A2)  at ( 7, 0 ){$A_2$};

    \node (Y)   at ( 10, 0 ) {$Y_{2}$};
\end{scope}
\begin{scope}[>={Stealth[black]},
              every node/.style={fill=white,circle},
              every edge/.style={draw=black,very thick}]
    
    \path [->] (V1)  edge (A1);
    \path [->] (V1)  edge (A2);
    
    \path [->] (V2)  edge (A2);
    
    \path [->] (V1)[bend right=15]  edge (V2);

    \path [->] (Z)  edge (ZA);
    \path [->] (Z)  edge (ZY);

    \path [->] (ZY)  edge[bend left=15] (Y);

    \path [->] (ZA)  edge (A1);
    \path [->] (ZA)  edge[bend right=15] (A2);

    \path [->] (ZA)  edge (V2);
    
    \path [->] (A1)  edge (V2);
    
    \path [->] (A1)  edge (A2);
    
    \path [->] (A1)[bend right=25]  edge (Y);
    \path [->] (A2)  edge (Y);
\end{scope}
\end{tikzpicture}
}
\caption{Causal DAG representing a setting in which the $Z_{A}$ component of a modified treatment exerts an effect on $Y_{2}$ only through adherence, mediated by non-prognostic covariates, $V_{2}$, and the $Z_{Y}$ component exerts no effect on adherence.}
\label{fig: nonprognostic_side_effects_example_graphs}
\end{figure}
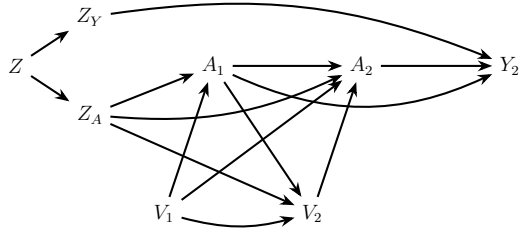

%% file: Extra/Figures/prognostic_side_effects_example_graphs.tex
\begin{figure}
    \centering
\scalebox{0.65}{
\begin{tikzpicture}
\begin{scope}[every node/.style={thick,draw=none}]


    \node (L1)   at ( 3, -3 ) {$L_1$};
    \node (L2)  at ( 6, -3 ) {$L_2$};

    \node (Z)   at ( 0, 0 ) {$Z$};
    \node (ZY)   at ( 1.5, 1 ) {$Z_Y$};
    \node (ZA)   at ( 1.5, -1 ) {$Z_A$};

    \node (A1)   at ( 4, 0 ) {$A_1$};
    \node (A2)  at ( 7, 0 ){$A_2$};

    \node (Y)   at ( 10, 0 ) {$Y_{2}$};
\end{scope}
\begin{scope}[>={Stealth[black]},
              every node/.style={fill=white,circle},
              every edge/.style={draw=black,very thick}]

    
    \path [->] (L1)  edge (A1);
    \path [->] (L1)  edge (A2);
    
    \path [->] (L2)  edge (A2);
    
    \path [->] (L1)[bend right=15]  edge (L2);

    \path [->] (L1)  edge (Y);
    \path [->] (L2)  edge (Y);

    \path [->] (Z)  edge (ZA);
    \path [->] (Z)  edge (ZY);

    \path [->] (ZY)  edge[bend left=15] (Y);

    \path [->] (ZA)  edge (A1);
    \path [->] (ZA)  edge[bend right=15] (A2);

    \path [->] (ZA)  edge (L2);
    
    \path [->] (A1)  edge (L2);
    
    \path [->] (A1)  edge (A2);
    
    \path [->] (A1)[bend right=25]  edge (Y);
    \path [->] (A2)  edge (Y);
\end{scope}
\end{tikzpicture}
}
\caption{Causal DAG representing a setting in which the $Z_{A}$ component of a modified treatment exerts an effect on $Y_{2}$ through adherence and prognostic covariates, $L_{2}$, and the $Z_{Y}$ component exerts no effect on adherence.}
\label{fig: prognostic_side_effects_example_graphs}
\end{figure}
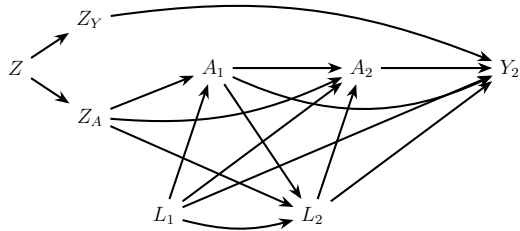

%% file: Extra/Figures/prognostic_side_effects_example_graphs2.tex
\begin{figure}
    \centering
\scalebox{0.65}{
\begin{tikzpicture}
\begin{scope}[every node/.style={thick,draw=none}]


    \node (L1)   at ( 3, 3 ) {$L_1$};
    \node (L2)  at ( 6, 3 ) {$L_2$};

    \node (Z)   at ( 0, 0 ) {$Z$};
    \node (ZY)   at ( 1.5, 1 ) {$Z_Y$};
    \node (ZA)   at ( 1.5, -1 ) {$Z_A$};

    \node (A1)   at ( 4, 0 ) {$A_1$};
    \node (A2)  at ( 7, 0 ){$A_2$};

    \node (Y)   at ( 10, 0 ) {$Y_{2}$};
\end{scope}
\begin{scope}[>={Stealth[black]},
              every node/.style={fill=white,circle},
              every edge/.style={draw=black,very thick}]

    
    \path [->] (L1)  edge (A1);
    \path [->] (L1)  edge (A2);
    
    \path [->] (L2)  edge (A2);
    
    \path [->] (L1)[bend left=15]  edge (L2);

    \path [->] (L1)  edge (Y);
    \path [->] (L2)  edge (Y);

    \path [->] (Z)  edge (ZA);
    \path [->] (Z)  edge (ZY);

    \path [->] (ZY)  edge (L2);

    \path [->] (ZY)  edge[bend left=15] (Y);

    \path [->] (ZA)  edge (A1);
    \path [->] (ZA)  edge[bend right=15] (A2);
    
    \path [->] (A1)  edge (L2);
    
    \path [->] (A1)  edge (A2);
    
    \path [->] (A1)[bend right=25]  edge (Y);
    \path [->] (A2)  edge (Y);
\end{scope}
\end{tikzpicture}
}
\caption{Causal DAG representing a setting in which the $Z_{A}$ component of a modified treatment exerts an effect on $Y_{2}$ only through adherence, while the $Z_{Y}$ component exerts an effect on adherence through its effect on $L_{2}$.}
\label{fig: prognostic_side_effects_example_graphs2}
\end{figure}
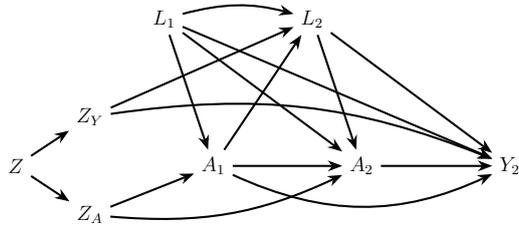

%% file: Extra/Figures/subset_prognostic_covariates.tex
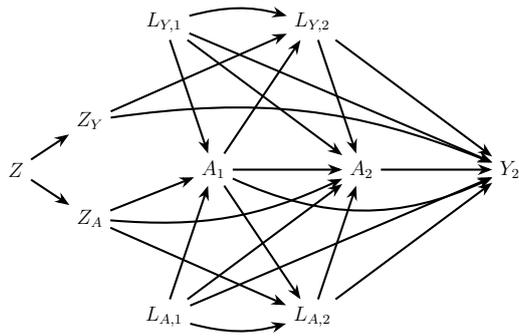
\begin{figure}
     \centering
\scalebox{0.65}{
\begin{tikzpicture}
\begin{scope}[every node/.style={thick,draw=none}]
        
    \node (LY1)   at ( 3, 3 ) {$L_{Y,1}$};
    \node (LY2)  at ( 6, 3 ) {$L_{Y,2}$};

    \node (LA1)   at ( 3, -3 ) {$L_{A,1}$};
    \node (LA2)  at ( 6, -3 ) {$L_{A,2}$};


    \node (Z)   at ( 0, 0 ) {$Z$};
    \node (ZY)   at ( 1.5, 1 ) {$Z_Y$};
    \node (ZA)   at ( 1.5, -1 ) {$Z_A$};

    \node (A1)   at ( 4, 0 ) {$A_1$};
    \node (A2)  at ( 7, 0 ){$A_2$};

    \node (Y)   at ( 10, 0 ) {$Y_{2}$};
\end{scope}
\begin{scope}[>={Stealth[black]},
              every node/.style={fill=white,circle},
              every edge/.style={draw=black,very thick}]


    
    \path [->] (LY1)  edge (A1);
    \path [->] (LY1)  edge (A2);
    
    \path [->] (LY2)  edge (A2);
    
    \path [->] (LY1)[bend left=15]  edge (LY2);

    \path [->] (LA1)  edge (A1);
    \path [->] (LA1)  edge (A2);
    
    \path [->] (LA2)  edge (A2);
    
    \path [->] (LA1)[bend right=15]  edge (LA2);

    \path [->] (LA1)  edge (Y);
    \path [->] (LA2)  edge (Y);

    \path [->] (LY1)  edge (Y);
    \path [->] (LY2)  edge (Y);

    \path [->] (Z)  edge (ZA);
    \path [->] (Z)  edge (ZY);

    \path [->] (ZY)  edge[bend left=15] (Y);

    \path [->] (ZY)  edge (LY2);

    \path [->] (ZA)  edge (A1);
    \path [->] (ZA)  edge[bend right=15] (A2);

    \path [->] (ZA)  edge (LA2);
    
    \path [->] (A1)  edge (LY2);

    \path [->] (A1)  edge (LA2);
    
    \path [->] (A1)  edge (A2);
    
    \path [->] (A1)[bend right=25]  edge (Y);
    \path [->] (A2)  edge (Y);
\end{scope}
\end{tikzpicture}
}
\caption{Causal DAG representing a setting in which both $Z_{Y}$ and $Z_{A}$ exert effects on adherence through different sets of prognostic covariates. The $Z_{A}$ component exerts an effect on $Y_{2}$ through adherence and $L_{A,2}$, while the $Z_{Y}$ component exerts an effect on adherence through its effect on $L_{Y,2}$.}
\label{fig: subset_prognostic_covariates}
\end{figure}

%% file: Extra/Figures/prognostic_side_effects_example_graphs_not_identified.tex
\begin{figure}
    \centering
\begin{subfigure}[b]{0.45\textwidth}
\scalebox{0.65}{
\begin{tikzpicture}
\begin{scope}[every node/.style={thick,draw=none}]

    \node (U) at (5.5, -5) {$U$};

    \node (L1)   at ( 3, -3 ) {$L_{A,1}$};
    \node (L2)  at ( 6, -3 ) {$L_{A,2}$};

    \node (Z)   at ( 0, 0 ) {$Z$};
    \node (ZY)   at ( 1.5, 1 ) {$Z_Y$};
    \node (ZA)   at ( 1.5, -1 ) {$Z_A$};

    \node (A1)   at ( 4, 0 ) {$A_1$};
    \node (A2)  at ( 7, 0 ){$A_2$};

    \node (Y)   at ( 10, 0 ) {$Y_{2}$};
\end{scope}
\begin{scope}[>={Stealth[black]},
              every node/.style={fill=white,circle},
              every edge/.style={draw=black,very thick}]

    \path [->] (U) edge (L2);
    \path [->] (U)[bend right = 15] edge (Y);
    
    \path [->] (L1)  edge (A1);
    \path [->] (L1)  edge (A2);
    
    \path [->] (L2)  edge (A2);
    
    \path [->] (L1)[bend right=15]  edge (L2);

    \path [->] (L1)  edge (Y);
    \path [->] (L2)  edge (Y);

    \path [->] (Z)  edge (ZA);
    \path [->] (Z)  edge (ZY);

    \path [->] (ZY)  edge[bend left=15] (Y);

    \path [->] (ZA)  edge (A1);
    \path [->] (ZA)  edge[bend right=15] (A2);

    \path [->] (ZA)  edge (L2);
    
    \path [->] (A1)  edge (L2);
    
    \path [->] (A1)  edge (A2);
    
    \path [->] (A1)[bend right=25]  edge (Y);
    \path [->] (A2)  edge (Y);
\end{scope}
\end{tikzpicture}
}
\caption{DAG1}
\end{subfigure}
\vfill
\begin{subfigure}[b]{0.45\textwidth}
\scalebox{0.65}{
\begin{tikzpicture}
\begin{scope}[every node/.style={thick,draw=none}]

    \node (U) at (4, -5) {$U$};

    \node (L1)   at ( 3, -3 ) {$L_{A,1}$};
    \node (L2)  at ( 6, -3 )  {$L_{A,2}$};

    \node (Z)   at ( 0, 0 ) {$Z$};
    \node (ZY)   at ( 1.5, 1 ) {$Z_Y$};
    \node (ZA)   at ( 1.5, -1 ) {$Z_A$};

    \node (A1)   at ( 4, 0 ) {$A_1$};
    \node (A2)  at ( 7, 0 ){$A_2$};

    \node (Y)   at ( 10, 0 ) {$Y_{2}$};
\end{scope}
\begin{scope}[>={Stealth[black]},
              every node/.style={fill=white,circle},
              every edge/.style={draw=black,very thick}]

    \path [->] (U)[bend right = 15] edge (Y);
    
    \path [->] (L1)  edge (A1);
    \path [->] (L1)  edge (A2);
    
    \path [->] (L2)  edge (A2);
    
    \path [->] (L1)[bend right=15]  edge (L2);

    \path [->] (L1)  edge (Y);
    \path [->] (L2)  edge (Y);

    \path [->] (Z)  edge (ZA);
    \path [->] (Z)  edge (ZY);

    \path [->] (ZY)  edge[bend left=15] (Y);

    \path [->] (ZA)  edge (A1);
    \path [->] (ZA)  edge[bend right=15] (A2);
    \path [->] (ZA)  edge[bend right=15] (U);

    \path [->] (ZA)  edge (L2);
    
    \path [->] (A1)  edge (L2);
    
    \path [->] (A1)  edge (A2);
    
    \path [->] (A1)[bend right=25]  edge (Y);
    \path [->] (A2)  edge (Y);
\end{scope}
\end{tikzpicture}
}
\caption{DAG2}
\end{subfigure}
\hfill
\begin{subfigure}[b]{0.45\textwidth}
\scalebox{0.65}{
\begin{tikzpicture}
\begin{scope}[every node/.style={thick,draw=none}]

    \node (U) at (2, 0.5) {$U$};

    \node (L1)   at ( 3, -3 ) {$L_{A,1}$};
    \node (L2)  at ( 6, -3 )  {$L_{A,2}$};

    \node (Z)   at ( 0, 0 ) {$Z$};
    \node (ZY)   at ( 1, 2 ) {$Z_Y$};
    \node (ZA)   at ( 1.5, -1 ) {$Z_A$};

    \node (A1)   at ( 4, 0 ) {$A_1$};
    \node (A2)  at ( 7, 0 ){$A_2$};

    \node (Y)   at ( 10, 0 ) {$Y_{2}$};
\end{scope}
\begin{scope}[>={Stealth[black]},
              every node/.style={fill=white,circle},
              every edge/.style={draw=black,very thick}]
    
    \path [->] (L1)  edge (A1);
    \path [->] (L1)  edge (A2);
    
    \path [->] (L2)  edge (A2);
    
    \path [->] (L1)[bend right=15]  edge (L2);

    \path [->] (L1)  edge (Y);
    \path [->] (L2)  edge (Y);

    \path [->] (Z)  edge (ZA);
    \path [->] (Z)  edge (ZY);

    \path [->] (ZY)  edge[bend left=15] (Y);
    \path [->] (ZY)  edge (U);
    \path [->] (U)  edge[bend right=10] (L2);

    \path [->] (ZA)  edge (A1);
    \path [->] (ZA)  edge[bend right=15] (A2);

    \path [->] (ZA)  edge (L2);
    
    \path [->] (A1)  edge (L2);
    
    \path [->] (A1)  edge (A2);
    
    \path [->] (A1)[bend right=25]  edge (Y);
    \path [->] (A2)  edge (Y);
\end{scope}
\end{tikzpicture}
}
\caption{DAG3}
\end{subfigure}
\caption{Causal DAGs representing settings where independence conditions required for separable effects do not hold due to an unmeasured variable, $U$.}
\label{fig: prognostic_side_effects_example_graphs_not_identified}
\end{figure}
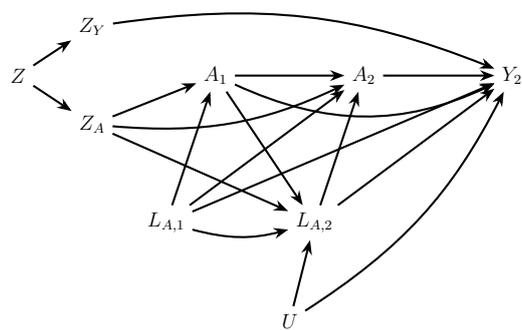
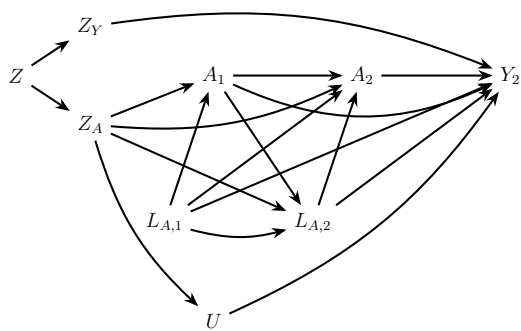
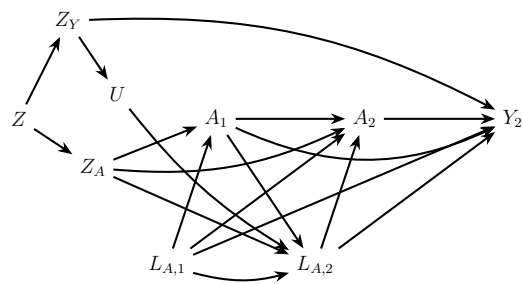

%% file: Body/appendix.tex
\appendix
\begin{center}
\textbf{\large Supplemental Materials for ``Separable effects for adherence''}
\end{center}
\setcounter{equation}{0}
\setcounter{figure}{0}
\setcounter{table}{0}
\setcounter{page}{1}
\setcounter{section}{0}
\makeatletter
\renewcommand{\theequation}{S\arabic{equation}}
\renewcommand{\thefigure}{S\arabic{figure}}
\renewcommand{\thetable}{S\arabic{table}}
\renewcommand{\bibnumfmt}[1]{[S#1]}
\renewcommand{\thesection}{S\arabic{section}}

\section{Equivalence of the g-formula and the inverse probability weighted expression}
\label{section: IPW proof}

\noindent In this section we prove the equivalence of the g-formula and its inverse probability weighted expression. The weighted representation is 

\begin{align}
    \sum_{k=1}^{K} \lambda_{Y,k} \prod_{j=1}^{k-1} [1-\lambda_{Y,j}]
    \label{IPW_1}
\end{align}

\noindent where

\begin{align}
    \lambda_{Y,k} = \frac{\mathrm{E}[Y_k(1-Y_{k-1})W_{Y,k} \mid Z=z_{A}]}{\mathrm{E}[(1-Y_{k-1})W_{Y,k} \mid Z=z_{A}]}
    \label{IPW_2}
\end{align}

\noindent with weights, $W_{Y,k}$, defined as

\begin{align*}
W_{Y,k}= & \prod_{j=1}^{k}\frac{f(Y_{j} \mid \overline{Y}_{j-1}=0, \overline{L}_{Y,j},\overline{L}_{A,j},\overline{A}_{j}, Z=z_{Y})}{f(Y_{j} \mid \overline{Y}_{j-1}=0, \overline{L}_{Y,j},\overline{L}_{A,j},\overline{A}_{j},Z=z_{A})} \times \\
& \prod_{j=1}^{k} \Bigg[\frac{\Pr(Z=z_{Y} \mid \overline{Y}_{j-1}=0,\overline{L}_{Y,j},\overline{L}_{A,j},\overline{A}_{j-1})}{\Pr(Z=z_{A} \mid \overline{Y}_{j-1}=0,\overline{L}_{Y,j},\overline{L}_{A,j},\overline{A}_{j-1})} \times \\
& \hspace{1cm} \frac{\Pr(Z=z_{A} \mid \overline{Y}_{j-1}=0,\overline{L}_{Y,j-1},\overline{L}_{A,j},\overline{A}_{j-1})}{\Pr(Z=z_{Y} \mid \overline{Y}_{j-1}=0,\overline{L}_{Y,j-1},\overline{L}_{A,j},\overline{A}_{j-1})}\Bigg],
\end{align*}

which can be reexpressed in the following way,

\begin{align*}
W_{Y,k} &= \prod_{j=1}^{k}\frac{f(Y_{j} \mid \overline{Y}_{j-1}=0, \overline{L}_{Y,j},\overline{L}_{A,j},\overline{A}_{j}, Z=z_{Y})}{f(Y_{j} \mid \overline{Y}_{j-1}=0, \overline{L}_{Y,j},\overline{L}_{A,j},\overline{A}_{j},Z=z_{A})} \times \\
& \hspace{0.5cm} \prod_{j=1}^{k} \Bigg[\frac{\frac{f(Z=z_{Y},L_{Y,j} \mid \overline{Y}_{j-1}=0,\overline{L}_{Y,j-1},\overline{L}_{A,j},\overline{A}_{j-1})}{f(L_{Y,j} \mid \overline{Y}_{j-1}=0,\overline{L}_{Y,j-1},\overline{L}_{A,j},\overline{A}_{j-1})}}{\frac{f(Z=z_{A},L_{Y,j} \mid \overline{Y}_{j-1}=0,\overline{L}_{Y,j-1},\overline{L}_{A,j},\overline{A}_{j-1})}{f(L_{Y,j} \mid \overline{Y}_{j-1}=0,\overline{L}_{Y,j-1},\overline{L}_{A,j},\overline{A}_{j-1})}} \times \\
& \hspace{1.5cm} \frac{\Pr(Z=z_{A} \mid \overline{Y}_{j-1}=0,\overline{L}_{Y,j-1},\overline{L}_{A,j},\overline{A}_{j-1})}{\Pr(Z=z_{Y} \mid \overline{Y}_{j-1}=0,\overline{L}_{Y,j-1},\overline{L}_{A,j},\overline{A}_{j-1})}\Bigg] \\
&= \prod_{j=1}^{k}\frac{f(Y_{j} \mid \overline{Y}_{j-1}=0, \overline{L}_{Y,j},\overline{L}_{A,j},\overline{A}_{j}, Z=z_{Y})}{f(Y_{j} \mid \overline{Y}_{j-1}=0, \overline{L}_{Y,j},\overline{L}_{A,j},\overline{A}_{j},Z=z_{A})} \times \\
& \hspace{0.5cm} \prod_{j=1}^{k} \frac{f(L_{Y,j} \mid \overline{Y}_{j-1}=0,\overline{L}_{Y,j-1},\overline{L}_{A,j},\overline{A}_{j-1},Z=z_{Y})}{f(L_{Y,j} \mid \overline{Y}_{j-1}=0,\overline{L}_{Y,j-1},\overline{L}_{A,j},\overline{A}_{j-1},Z=z_{A})}. \\
\end{align*}

First, we use \eqref{IPW_2} to rewrite \eqref{IPW_1} as

\begin{align*}
    \sum_{k=1}^{K} &\lambda_{Y,k} \prod_{j=1}^{k-1} [1-\lambda_{Y,j}] \\
    = &\sum_{k=1}^{K} \mathrm{E}[Y_{k}(1-Y_{k-1})W_{Y,k} \mid Z=z_{A}] \times \\
    & \hspace{1cm} \prod_{j=1}^{k} \frac{\mathrm{E}[(1-Y_{j-2})W_{Y,j-1} \mid Z=z_{A}] - \mathrm{E}[Y_{j-1}(1-Y_{j-2})W_{Y,j-1} \mid Z=z_{A}]}{\mathrm{E}[(1-Y_{j-1})W_{Y,j} \mid Z=z_{A}]} \\
    = &\sum_{k=1}^{K} \mathrm{E}[Y_{k}(1-Y_{k-1})W_{Y,k} \mid Z=z_{A}],
\end{align*}

\noindent where any variables with $k \leq 0$ are set to an arbitrary constant by convention, and in the second step we use that 

\begin{align*}
\mathrm{E}[(1-Y_{j-1})W_{Y,j} \mid Z=z_{A}] = \mathrm{E}[(1-Y_{j-2})W_{Y,j-1} \mid Z=z_{A}] - \mathrm{E}[Y_{j-1}(1-Y_{j-2})W_{Y,j-1} \mid Z=z_{A}]
\end{align*}

\noindent because, using the definition of an expectation,

\begin{align*}
\mathrm{E}&[(1-Y_{j-1})W_{Y,j} \mid Z=z_{A}] \\
&= \sum_{\overline{l}_{j},\overline{a}_{j},y_{j}} W_{Y,j} f(y_{j}, \overline{l}_{j}, \overline{a}_{j}, \overline{Y}_{j-1} = 0 \mid Z=z_{A}) \\
&= \sum_{\overline{l}_{j-1},\overline{a}_{j-1}} \prod_{s=1}^{j-1} \Bigg[\frac{f(y_{s} \mid \overline{Y}_{s-1}=0,\overline{l}_{Y,s},\overline{l}_{A,s},\overline{a}_{s}, Z=z_{Y})}{f(y_{s} \mid \overline{Y}_{s-1}=0,\overline{l}_{Y,s},\overline{l}_{A,s},\overline{a}_{s}, Z=z_{A})} \times \\
& \hspace{2.5cm} \frac{f(l_{Y,s} \mid \overline{Y}_{s-1}=0,\overline{l}_{Y,s-1},\overline{l}_{A,s},\overline{a}_{s-1}, Z=z_{Y})}{f(l_{Y,s} \mid \overline{Y}_{s-1}=0,\overline{l}_{Y,s-1},\overline{l}_{A,s},\overline{a}_{s-1}, Z=z_{A})}\Bigg]
\times \\
&\hspace{1cm} \sum_{l_{j},a_{j},y_{j}} \Bigg[\frac{f(y_{j} \mid \overline{Y}_{j-1}=0,\overline{l}_{Y,j},\overline{l}_{A,j},\overline{a}_{j}, Z=z_{Y})}{f(y_{j} \mid \overline{Y}_{j-1}=0,\overline{l}_{Y,j},\overline{l}_{A,j},\overline{a}_{j}, Z=z_{A})} \times \\
& \hspace{2cm} \frac{f(l_{Y,j} \mid \overline{Y}_{j-1}=0,\overline{l}_{Y,j-1},\overline{l}_{A,j},\overline{a}_{j-1}, Z=z_{Y})}{f(l_{Y,j} \mid \overline{Y}_{j-1}=0,\overline{l}_{Y,j-1},\overline{l}_{A,j},\overline{a}_{j-1}, Z=z_{A})}\Bigg] f(y_{j}, \overline{l}_{j}, \overline{a}_{j}, \overline{Y}_{j-1} = 0 \mid Z=z_{A}) \\
&= \sum_{\overline{l}_{j-1},\overline{a}_{j-1}} W_{Y,j-1} f(\overline{l}_{j-1}, \overline{a}_{j-1}, \overline{Y}_{j-1}=0 \mid Z=z_{A}) \\
&= \mathrm{E}[(1-Y_{j-1})W_{Y,j-1} \mid Z=z_{A}]
\end{align*}

\noindent and, using linearity of expectations and that $Y_{j-1}=0$ implies $\overline{Y}_{j-2}=0$,

\begin{align*}
\mathrm{E}[(1-Y_{j-1})W_{Y,j-1} \mid Z=z_{A}] &= \mathrm{E}[(1-Y_{j-1})(1-Y_{j-2})W_{Y,j-1} \mid Z=z_{A}] \\
&= \mathrm{E}[(1-Y_{j-2})W_{Y,j-1} - Y_{j-1}(1-Y_{j-2})W_{Y,j-1} \mid Z=z_{A}] \\
&= \mathrm{E}[(1-Y_{j-2})W_{Y,j-1} \mid Z=z_{A}] - \mathrm{E}[Y_{j-1}(1-Y_{j-2})W_{Y,j-1} \mid Z=z_{A}].
\end{align*}

With the reexpressed weighted representation, we use the definition of an expectation to write

\begin{align*}
&\sum_{k=1}^{K}\mathrm{E}[Y_{k}(1-Y_{k-1})W_{Y,k} \mid Z=z_{A}] \\
&= \sum_{k=1}^{K}\sum_{\overline{l}_{k},\overline{a}_{k},\overline{y}_{k}} y_{k}(1-y_{k-1})W_{Y,k}f(\overline{l}_{k},\overline{a}_{k},\overline{y}_{k} \mid Z=z_{A})\\
&= \sum_{k=1}^{K}\sum_{\overline{l}_{k},\overline{a}_{k},y_{k}} y_{k} W_{Y,k}f(y_{k},\overline{Y}_{k-1}=0,\overline{l}_{k},\overline{a}_{k} \mid Z=z_{A})\\
&= \sum_{k=1}^{K}\sum_{\overline{l}_{k},\overline{a}_{k},y_{k}} y_{k} W_{Y,k} f(y_{k} \mid \overline{Y}_{k-1}=0,\overline{l}_{k},\overline{a}_{k}, Z=z_{A}) f(a_{k} \mid \overline{Y}_{k-1}=0,\overline{l}_{k},\overline{a}_{k-1}, Z=z_{A}) \times \\
& \hspace{2cm} f(l_{k} \mid \overline{Y}_{k-1}=0,\overline{l}_{k-1},\overline{a}_{k-1}, Z=z_{A}) f(\overline{Y}_{k-1}=0,\overline{l}_{k-1},\overline{a}_{k-1} \mid Z=z_{A})\\
&= \sum_{k=1}^{K}\sum_{\overline{l}_{k},\overline{a}_{k},y_{k}} y_{k} W_{Y,k} f(y_{k} \mid \overline{Y}_{k-1}=0,\overline{l}_{k},\overline{a}_{k}, Z=z_{A}) \times \\
& \hspace{1cm} \Bigg[ \prod_{j=1}^{k} \Pr(Y_{j-1}=0 \mid \overline{Y}_{j-2}=0, \overline{l}_{j-1},\overline{a}_{j-1}, Z=z_{A}) \times \\
& \hspace{2cm} f(l_{j} \mid \overline{Y}_{j-1}=0,\overline{l}_{j-1},\overline{a}_{j-1}, Z=z_{A}) f(a_{j} \mid \overline{Y}_{j-1}=0,\overline{l}_{j},\overline{a}_{j-1}, Z=z_{A}) \Bigg] \\
&= \sum_{k=1}^{K}\sum_{\overline{l}_{k},\overline{a}_{k},y_{k}} y_{k} W_{Y,k} f(y_{k} \mid \overline{Y}_{k-1}=0,\overline{l}_{Y,k},\overline{l}_{A,k},\overline{a}_{k}, Z=z_{A}) \times \\
& \Bigg[ \prod_{j=1}^{k} \Pr(Y_{j-1}=0 \mid \overline{Y}_{j-2}=0, \overline{l}_{Y,j-1},\overline{l}_{A,j-1},\overline{a}_{j-1}, Z=z_{A}) f(l_{Y,j} \mid \overline{Y}_{j-1}=0,\overline{l}_{Y,j-1},\overline{l}_{A,j},\overline{a}_{j-1}, Z=z_{A}) \times \\
& \ f(l_{A,j} \mid \overline{Y}_{j-1}=0,\overline{l}_{Y,j-1},\overline{l}_{A,j-1},\overline{a}_{j-1}, Z=z_{A}) f(a_{j} \mid \overline{Y}_{j-1}=0,\overline{l}_{Y,j},\overline{l}_{A,j},\overline{a}_{j-1}, Z=z_{A}) \Bigg], \\
\end{align*}

\noindent where we use the definition of conditional probability and that $L_{j} = (L_{Y,j},L_{A,j})$ in the last two steps. Then, plugging in the expression for $W_{Y,k}$, we have

\begin{align*}
&\sum_{k=1}^{K}\sum_{\overline{l}_{k},\overline{a}_{k}} \Pr(Y_{k}=1 \mid \overline{Y}_{k-1}=0,\overline{l}_{Y,k},\overline{l}_{A,k},\overline{a}_{k}, Z=z_{Y}) \times \\
& \Bigg[ \prod_{j=1}^{k} \Pr(Y_{j-1}=0 \mid \overline{Y}_{j-2}=0, \overline{l}_{Y,j-1},\overline{l}_{A,j-1},\overline{a}_{j-1}, Z=z_{Y}) f(l_{Y,j} \mid \overline{Y}_{j-1}=0,\overline{l}_{Y,j-1},\overline{l}_{A,j},\overline{a}_{j-1}, Z=z_{Y}) \times \\
& \ f(l_{A,j} \mid \overline{Y}_{j-1}=0,\overline{l}_{Y,j-1},\overline{l}_{A,j-1},\overline{a}_{j-1}, Z=z_{A}) f(a_{j} \mid \overline{Y}_{j-1}=0,\overline{l}_{Y,j},\overline{l}_{A,j},\overline{a}_{j-1}, Z=z_{A}) \Bigg], \\
\end{align*}

\noindent which is the g-formula for the separable effect of $Z_{Y}$ under $Z_{A}=z_{A}$.

\section{An alternative inverse probability weighted expression}
\label{section: alternative_IPW}

\noindent In Supplementary Materials Section~\ref{section: IPW proof}, we considered one inverse probability weighted representation of the g-formula for the separable effect of $Z_{Y}$ under $Z_{A}=z_{A}$. In this section, we consider an alternative expression characterized by defining

\begin{align*}
    \lambda_{A,k} = \frac{\mathrm{E}[Y_k(1-Y_{k-1})W_{A,k} \mid Z=z_{Y}]}{\mathrm{E}[(1-Y_{k-1})W_{A,k} \mid Z=z_{Y}]} 
\end{align*}

\noindent and weights, $W_{A,k}$, defined as

\begin{align*}
W_{A,k}= \prod_{j=1}^{k} \Bigg[ &\frac{f(A_{j} \mid \overline{Y}_{j-1}=0,\overline{L}_{Y,j},\overline{L}_{A,j},\overline{A}_{j-1}, Z=z_{A})}{f(A_{j} \mid \overline{Y}_{j-1}=0,\overline{L}_{Y,j},\overline{L}_{A,j},\overline{A}_{j-1}, Z=z_{Y})} \times \\
& \frac{\Pr(Z=z_{A} \mid \overline{Y}_{j-1}=0,\overline{L}_{Y,j-1},\overline{L}_{A,j},\overline{A}_{j-1})}{\Pr(Z=z_{Y} \mid \overline{Y}_{j-1}=0,\overline{L}_{Y,j-1},\overline{L}_{A,j},\overline{A}_{j-1})} \times \\
& \frac{\Pr(Z=z_{Y} \mid \overline{Y}_{j-1}=0,\overline{L}_{Y,j-1},\overline{L}_{A,j-1},\overline{A}_{j-1})}{\Pr(Z=z_{A} \mid \overline{Y}_{j-1}=0,\overline{L}_{Y,j-1},\overline{L}_{A,j-1},\overline{A}_{j-1})}\Bigg]
\end{align*}

which can be reexpressed as

\begin{align*}
W_{A,k} = \prod_{j=1}^{k} \Bigg[ &\frac{f(A_{j} \mid \overline{Y}_{j-1}=0,\overline{L}_{Y,j},\overline{L}_{A,j},\overline{A}_{j-1}, Z=z_{A})}{f(A_{j} \mid \overline{Y}_{j-1}=0,\overline{L}_{Y,j},\overline{L}_{A,j},\overline{A}_{j-1}, Z=z_{Y})} \times \\
& \frac{\frac{f(Z=z_{A}, L_{A,j} \mid \overline{Y}_{j-1}=0,\overline{L}_{Y,j-1},\overline{L}_{A,j-1},\overline{A}_{j-1})}{f(L_{A,j} \mid \overline{Y}_{j-1}=0,\overline{L}_{Y,j-1},\overline{L}_{A,j-1},\overline{A}_{j-1})}}{\frac{f(Z=z_{Y}, L_{A,j} \mid \overline{Y}_{j-1}=0,\overline{L}_{Y,j-1},\overline{L}_{A,j-1},\overline{A}_{j-1})}{f(L_{A,j} \mid \overline{Y}_{j-1}=0,\overline{L}_{Y,j-1},\overline{L}_{A,j-1},\overline{A}_{j-1})}} \times \\
& \frac{\Pr(Z=z_{Y} \mid \overline{Y}_{j-1}=0,\overline{L}_{Y,j-1},\overline{L}_{A,j-1},\overline{A}_{j-1})}{\Pr(Z=z_{A} \mid \overline{Y}_{j-1}=0,\overline{L}_{Y,j-1},\overline{L}_{A,j-1},\overline{A}_{j-1})}\Bigg] \\
= \prod_{j=1}^{k} \Bigg[ &\frac{f(A_{j} \mid \overline{Y}_{j-1}=0,\overline{L}_{Y,j},\overline{L}_{A,j},\overline{A}_{j-1}, Z=z_{A})}{f(A_{j} \mid \overline{Y}_{j-1}=0,\overline{L}_{Y,j},\overline{L}_{A,j},\overline{A}_{j-1}, Z=z_{Y})} \times \\
& \frac{f(L_{A,j} \mid \overline{Y}_{j-1}=0,\overline{L}_{Y,j-1},\overline{L}_{A,j-1},\overline{A}_{j-1}, Z=z_{A})}{f(L_{A,j} \mid \overline{Y}_{j-1}=0,\overline{L}_{Y,j-1},\overline{L}_{A,j-1},\overline{A}_{j-1}, Z=z_{Y})}\Bigg].
\end{align*}

\subsection{Equivalence of the g-formula and the alternative inverse probability weighted expression}
\label{section: alternative IPW proof}

\begin{align*}
&\sum_{k=1}^{K}\mathrm{E}[Y_{k}(1-Y_{k-1})W_{A,k} \mid Z=z_{Y}] \\
&= \sum_{k=1}^{K}\sum_{\overline{l}_{k},\overline{a}_{k},\overline{y}_{k}} y_{k}(1-y_{k-1})W_{A,k}f(\overline{l}_{k},\overline{a}_{k},\overline{y}_{k} \mid Z=z_{Y})\\
&= \sum_{k=1}^{K}\sum_{\overline{l}_{k},\overline{a}_{k}} W_{A,k}f(Y_{k}=1,\overline{Y}_{k-1}=0,\overline{l}_{k},\overline{a}_{k} \mid Z=z_{Y})\\
&= \sum_{k=1}^{K}\sum_{\overline{l}_{k},\overline{a}_{k}} W_{A,k} \Pr(Y_{k}=1 \mid \overline{Y}_{k-1}=0,\overline{l}_{k},\overline{a}_{k}, Z=z_{Y}) f(a_{k} \mid \overline{Y}_{k-1}=0,\overline{l}_{k},\overline{a}_{k-1}, Z=z_{Y}) \times \\
& \hspace{2cm} f(l_{k} \mid \overline{Y}_{k-1}=0,\overline{l}_{k-1},\overline{a}_{k-1}, Z=z_{Y}) f(\overline{Y}_{k-1}=0,\overline{l}_{k-1},\overline{a}_{k-1} \mid Z=z_{Y})\\
&= \sum_{k=1}^{K}\sum_{\overline{l}_{k},\overline{a}_{k}} W_{A,k} \Pr(Y_{k}=1 \mid \overline{Y}_{k-1}=0,\overline{l}_{k},\overline{a}_{k}, Z=z_{Y}) \times \\
& \hspace{1cm} \Bigg[ \prod_{j=1}^{k} \Pr(Y_{j-1}=0 \mid \overline{Y}_{j-2}=0, \overline{l}_{j-1},\overline{a}_{j-1}, Z=z_{Y}) \times \\
& \hspace{2cm} f(l_{j} \mid \overline{Y}_{j-1}=0,\overline{l}_{j-1},\overline{a}_{j-1}, Z=z_{Y}) f(a_{j} \mid \overline{Y}_{j-1}=0,\overline{l}_{j},\overline{a}_{j-1}, Z=z_{Y}) \Bigg] \\
&= \sum_{k=1}^{K}\sum_{\overline{l}_{k},\overline{a}_{k}} W_{A,k} \Pr(Y_{k}=1 \mid \overline{Y}_{k-1}=0,\overline{l}_{Y,k},\overline{l}_{A,k},\overline{a}_{k}, Z=z_{Y}) \times \\
& \Bigg[ \prod_{j=1}^{k} \Pr(Y_{j-1}=0 \mid \overline{Y}_{j-2}=0, \overline{l}_{Y,j-1},\overline{l}_{A,j-1},\overline{a}_{j-1}, Z=z_{Y}) f(l_{Y,j} \mid \overline{Y}_{j-1}=0,\overline{l}_{Y,j-1},\overline{l}_{A,j},\overline{a}_{j-1}, Z=z_{Y}) \times \\
& \ f(l_{A,j} \mid \overline{Y}_{j-1}=0,\overline{l}_{Y,j-1},\overline{l}_{A,j-1},\overline{a}_{j-1}, Z=z_{Y}) f(a_{j} \mid \overline{Y}_{j-1}=0,\overline{l}_{Y,j},\overline{l}_{A,j},\overline{a}_{j-1}, Z=z_{Y}) \Bigg], \\
\end{align*}

\noindent which is equivalent to the g-formula after the expression for $W_{A,k}$ is plugged in.

\subsection{Inverse probability weighted estimation}

\begin{footnotesize}

\begin{enumerate}
\label{algorithm: IPW2} 
    \item Using the entire data set, fit a pooled (over time) parametric regression model for 
    \begin{itemize}
    \item $f(A_{k}=a_{k} \mid \overline{Y}_{k-1}=0,\overline{l}_{Y,k},\overline{l}_{A,k},\overline{a}_{k-1}, Z)$,
    \item $\Pr(Z = z \mid \overline{Y}_{k-1}=0,\overline{l}_{Y,k-1},\overline{l}_{A,k},\overline{a}_{k-1})$, and
    \item $\Pr(Z = z \mid \overline{Y}_{k-1}=0,\overline{l}_{Y,k-1},\overline{l}_{A,k-1},\overline{a}_{k-1})$.
    \end{itemize}
    For example, we might assume pooled logistic regression models. 
    \item For each row in the data set for each individual with $Z=z_{A}$ at each time interval, $k$, in $1, \ldots, K$:
    \begin{enumerate}
        \item Obtain predicted values 
        \begin{itemize}
            \item $\widehat{f}(A_{k} \mid \overline{Y}_{k-1}=0,\overline{L}_{Y,k},\overline{L}_{A,k},\overline{A}_{k-1}, Z=z_{A})$,
            \item $\widehat{f}(A_{k} \mid \overline{Y}_{k-1}=0,\overline{L}_{Y,k},\overline{L}_{A,k},\overline{A}_{k-1}, Z=z_{Y})$,
            \item $\widehat{\Pr}(Z = z_{A} \mid \overline{Y}_{k-1}=0,\overline{L}_{Y,k-1},\overline{L}_{A,k},\overline{A}_{k-1})$,
            \item $\widehat{\Pr}(Z = z_{Y} \mid \overline{Y}_{k-1}=0,\overline{L}_{Y,k-1},\overline{L}_{A,k},\overline{A}_{k-1})$,
            \item $\widehat{\Pr}(Z = z_{A} \mid \overline{Y}_{k-1}=0,\overline{L}_{Y,k-1},\overline{L}_{A,k-1},\overline{A}_{k-1})$, and
            \item $\widehat{\Pr}(Z = z_{Y} \mid \overline{Y}_{k-1}=0,\overline{L}_{Y,k-1},\overline{L}_{A,k-1},\overline{A}_{k-1})$.
        \end{itemize}
        
        \item Evaluate 
        
        \begin{enumerate}
         \item 
         \begin{align*}
            \widehat{W}_{A,k}= \prod_{j=1}^{k} \Bigg[ &\frac{\widehat{f}(A_{k} \mid \overline{Y}_{j-1}=0,\overline{L}_{Y,j},\overline{L}_{A,j},\overline{A}_{j-1}, Z=z_{A})}{\widehat{f}(A_{k} \mid \overline{Y}_{j-1}=0,\overline{L}_{Y,j},\overline{L}_{A,j},\overline{A}_{j-1}, Z=z_{Y})} \times \\
            & \frac{\widehat{\Pr}(Z = z_{A} \mid \overline{Y}_{j-1}=0,\overline{L}_{Y,j-1},\overline{L}_{A,j},\overline{A}_{j-1})}{\widehat{\Pr}(Z = z_{Y} \mid \overline{Y}_{j-1}=0,\overline{L}_{Y,j-1},\overline{L}_{A,j},\overline{A}_{j-1})} \times \\
            & \frac{\widehat{\Pr}(Z = z_{Y} \mid \overline{Y}_{j-1}=0,\overline{L}_{Y,j-1},\overline{L}_{A,j-1},\overline{A}_{j-1})}{\widehat{\Pr}(Z = z_{A} \mid \overline{Y}_{j-1}=0,\overline{L}_{Y,j-1},\overline{L}_{A,j-1},\overline{A}_{j-1})}\Bigg]
        \end{align*}
        or, when $L_{k}=(L_{Y,k}, \emptyset), k=1,\ldots,K$, the simplified weight expression
        \item 
        \begin{align*}
            \widehat{W}_{A,k}= \prod_{j=1}^{k} \frac{\widehat{f}(A_{j} \mid \overline{Y}_{j-1}=0,\overline{L}_{Y,j},\overline{L}_{A,j},\overline{A}_{j-1}, Z=z_{A})}{\widehat{f}(A_{j} \mid \overline{Y}_{j-1}=0,\overline{L}_{Y,j},\overline{L}_{A,j},\overline{A}_{j-1}, Z=z_{Y})}.
        \end{align*}
        \end{enumerate}
    \end{enumerate}
    \item Compute the risk of failure by the end of $K$ as $\sum_{k=1}^{K} \widehat{\lambda}_{A,k} \displaystyle  \prod_{j=1}^{k-1} [1-\widehat{\lambda}_{A,j}]$ with the estimated weights, $\widehat{W}_{A,k}$, in $\widehat{\lambda}_{A,k} = \frac{\widehat{E}[Y_k(1-Y_{k-1})\widehat{W}_{A,k} \mid Z=z_{Y}]}{\widehat{E}[(1-Y_{k-1})\widehat{W}_{A,k} \mid Z=z_{Y}]}$.
\end{enumerate}

\end{footnotesize}

\section{Example using simulated data}
\label{section: toy_example}

\subsection{Design}

To illustrate an implementation of the separable effects for adherence, we simulated data from a hypothetical randomized trial in which investigators compare two different medication treatment strategies. Each individual is assigned to receive an ACEI ($Z=0$) or a thiazide diuretic ($Z=1$) for the duration of follow-up. Individuals are followed until death (the outcome) or the administrative end of the study (24 months following randomization). 

We generated data with $500,000$ individuals assigned to each treatment arm. We considered $L_{A,k}$ as an indicator of acute kidney injury (AKI), and $L_{Y,k}$ as an indicator of abnormal blood pressure in interval $k$. The parameters used to generate the data were inspired by literature on the impact of cost, AKI, and blood pressure on adherence to antihypertensive agents \citeS{baker2019peerS,mansfield2016prescriptionS,bidulka2020stoppingS,gupta2018impactS}. Covariates were generated among those surviving to interval $k$ according to the following models:

\begin{align*}
    L_{A,k} &\sim  \mathrm{Bernoulli}\left( 0.05+0.035*A_{k-1}-0.035*Z_{A}*A_{k-1} \right), \\
    L_{Y,k} &\sim  \mathrm{Bernoulli}\left( 0.95-0.60*A_{k-1}-0.15*Z_{Y}*A_{k-1} \right).   
\end{align*}

Adherence was generated among those surviving to interval $k$ using three different models, which are comparable to the causal models depicted in main text Figures~\ref{fig: cost_example_graphs}, \ref{fig: prognostic_side_effects_example_graphs}, and \ref{fig: subset_prognostic_covariates}, respectively:

\paragraph{Adherence model 1 (Figure~\ref{fig: adherence_model_1_dag})}

\begin{align*}
    A_{k} &\sim  \mathrm{Bernoulli}\left( 0.6+0.2*Z_{A}*A_{k-1} \right)
\end{align*}

\paragraph{Adherence model 2 (Figure~\ref{fig: adherence_model_2_dag})}

\begin{align*}
    A_{k} &\sim  \mathrm{Bernoulli}\left( 0.6+0.2*Z_{A}*A_{k-1}-0.5*L_{A,k} \right)
\end{align*}

\paragraph{Adherence model 3 (Figure~\ref{fig: adherence_model_3_dag})}

\begin{align*}
    A_{k} &\sim  \mathrm{Bernoulli}\left( 0.6+0.2*Z_{A}*A_{k-1}-0.5*L_{A,k}+0.2*L_{Y,k} \right)
\end{align*}

The outcome was generated among those surviving to interval $k$ according to the following model:

\begin{align*}
    Y_{k} &\sim  \mathrm{Bernoulli}\left( 0.035+0.01*L_{Y,k}-0.03*A_{k}+0.01*L_{A,k} \right)
\end{align*}

\input{Extra/Figures/adherence_model_1.tex}
\input{Extra/Figures/adherence_model_2.tex}
\input{Extra/Figures/adherence_model_3.tex}

In words, the data was generated with the risk of AKI and poorly controlled blood pressure being greater for individuals adherent to ACEIs compared to those adherent to thiazides. Adherence was generated using three different models: as a function of the $Z_{A}$ component of the initiated medication and past adherence (analogous to the main text example using medication cost); the $Z_{A}$ component of the initiated medication, past adherence, and AKI; and the $Z_{A}$ component of the initiated medication, past adherence, AKI, and poorly controlled blood pressure. The outcome risk was determined by adherence, poorly controlled blood pressure, and AKI incidence. Unlike Figures~\ref{fig: cost_example_graphs}, \ref{fig: prognostic_side_effects_example_graphs}, and \ref{fig: subset_prognostic_covariates}, the effect of the initiated medication on the outcome risk is entirely mediated through adherence and covariates. In other words, the data was generated as though there was no directed arrow from $Z_{Y}$ to $Y_{k}$ when $L_{Y,k}$ is included in the graphs (see Figures~\ref{fig: adherence_model_1_dag}, \ref{fig: adherence_model_2_dag}, and \ref{fig: adherence_model_3_dag}).

The code to reproduce the simulated data example is provided at: \url{https://github.com/KerollosWanis/separable_effects_for_adherence}.

\subsection{Simulation example results}

In Figures~\ref{fig: adherence_model_1_dag}, \ref{fig: adherence_model_2_dag}, and \ref{fig: adherence_model_3_dag} we present the distribution of the outcome, of adherence, and of the covariates (AKI and abnormal blood pressure) for the total (`intention-to-treat') effect comparison and the separable effect comparison. The latter compares outcomes under initiation of a modified version of ACEI to initiation of thiazide diuretics, with the adherence causing component of ACEI set to the value of the adherence causing component of thiazides (i.e. $\Pr[Y^{z_{A}=1, z_{Y}=0}_{k}=1]\mbox { vs. } \Pr[Y^{z_{A}=1, z_{Y}=1}_{k}=1]$). The separable effect was computed using the algorithm detailed in Section~\ref{section: estimation}.

To give intuition about how the interpretation of the separable effect varies for different data generating mechanisms and the causal graphs that represent them, we elaborate on the findings under each adherence model.

\subsubsection*{Adherence model 1: a causal structure where the separable effect balances adherence}

\begin{figure}[]
    \centering
    \includegraphics[scale = 0.75]{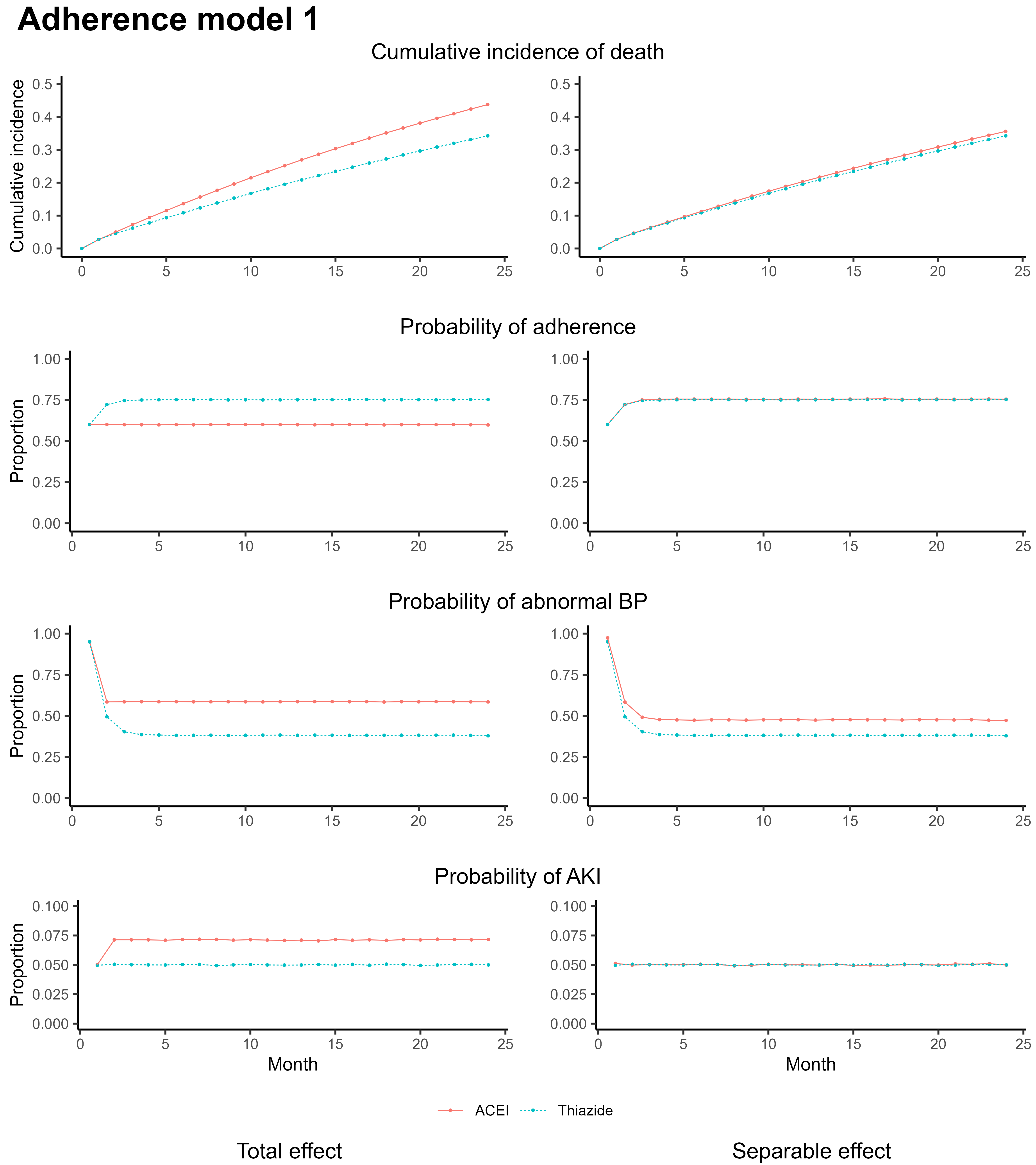}
    \caption{Cumulative incidence of death and probabilities of adherence, AKI, and abnormal blood pressure over 24 months of follow-up for a comparison of ACEI versus thiazide diuretic under the total (`intention-to-treat') effect in the left panels and under the separable effect (comparing a modified version of ACEI to thiazide diuretic) in the right panels. Data were generated using a simulated example with adherence depending only on the $Z_{A}$ component of the initiated medication and past adherence.}
    \label{adherence_model_1_results}
\end{figure}

In data generated using the first adherence model, only the $Z_{A}$ component of the initiated medication and past adherence exert effects on adherence. Figure~\ref{adherence_model_1_results} illustrates the difference in adherence under ACEI initiation versus thiazide initiation for the total effect comparison. As expected, consistent with the adherence model, adherence to thiazides is higher. Further, the probability of abnormal blood pressure is lower for thiazides, both because adherence to thiazides is higher and because adherence to thiazides is more effective than adherence to ACEIs. The probability of AKI is also lower for thiazides for identical reasons. 

Under the separable effect comparison, the probability of adherence is identical because $Z_{A}$ is set to the value it takes for thiazides. Further, the probability of AKI is identical because AKI depends only on the $Z_{A}$ component of the initiated medication and on past adherence. Lastly, the probability of abnormal blood pressure is more similar under the separable effect comparison than under the total effect comparison because the distribution of adherence is balanced. However, the probability of abnormal blood pressure is still lower for thiazides because the $Z_{Y}$ component exerts an effect on abnormal blood pressure causing thiazide adherence to be more effective than adherence to ACEIs.

\subsubsection*{Adherence model 2: another causal structure where the separable effect balances adherence}

In data generated using the second adherence model, the $Z_{A}$ component of the initiated medication, past adherence, and AKI exert effects on adherence. Under this data generating mechanism, the difference in adherence is even greater under ACEI initiation versus thiazide initiation because having an AKI reduces the probability of adherence and AKI is more likely for individuals adherent to an ACEI. But even though AKI exerts an effect on adherence, the separable effect still balances adherence because only $Z_{A}$, not $Z_{Y}$, affects the risk of AKI and, for both initiated medications, $Z_{A}$ is set to the value it takes for thiazides. As before, the probability of AKI is also identical for the separable effect because AKI depends only on the $Z_{A}$ component of the initiated medication and on past adherence.

\begin{figure}[]
    \centering
    \includegraphics[scale = 0.75]{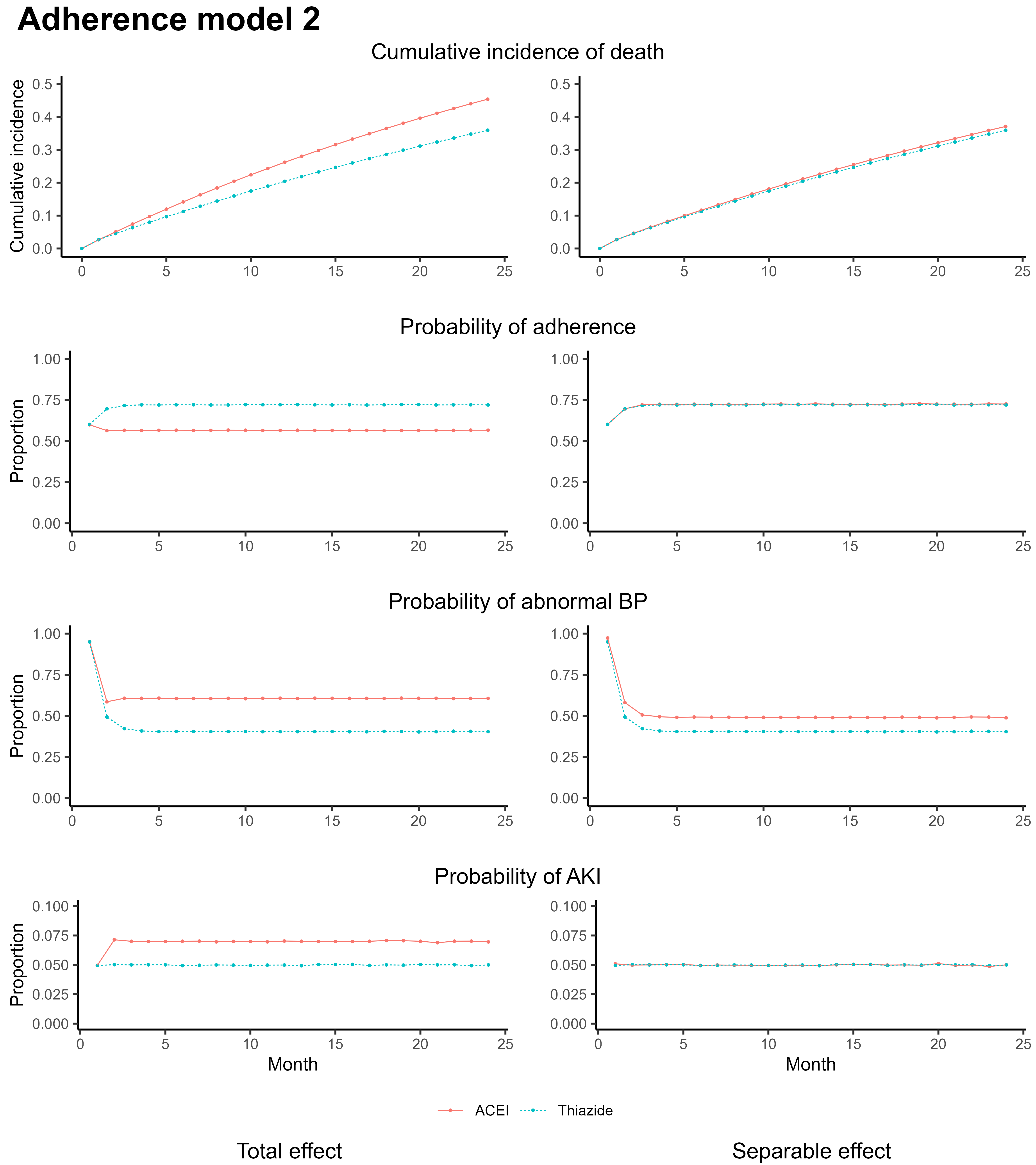}
    \caption{Cumulative incidence of death and probabilities of adherence, AKI, and abnormal blood pressure over 24 months of follow-up for a comparison of ACEI versus thiazide diuretic under the total (`intention-to-treat') effect in the left panels and under the separable effect (comparing a modified version of ACEI to thiazide diuretic) in the right panels. Data were generated using a simulated example with adherence depending only on the $Z_{A}$ component of the initiated medication, past adherence, and AKI.}
    \label{adherence_model_2_results}
\end{figure}

\subsubsection*{Adherence model 3: a causal structure where the separable effect does not balance adherence}

\begin{figure}[]
    \centering
    \includegraphics[scale = 0.75]{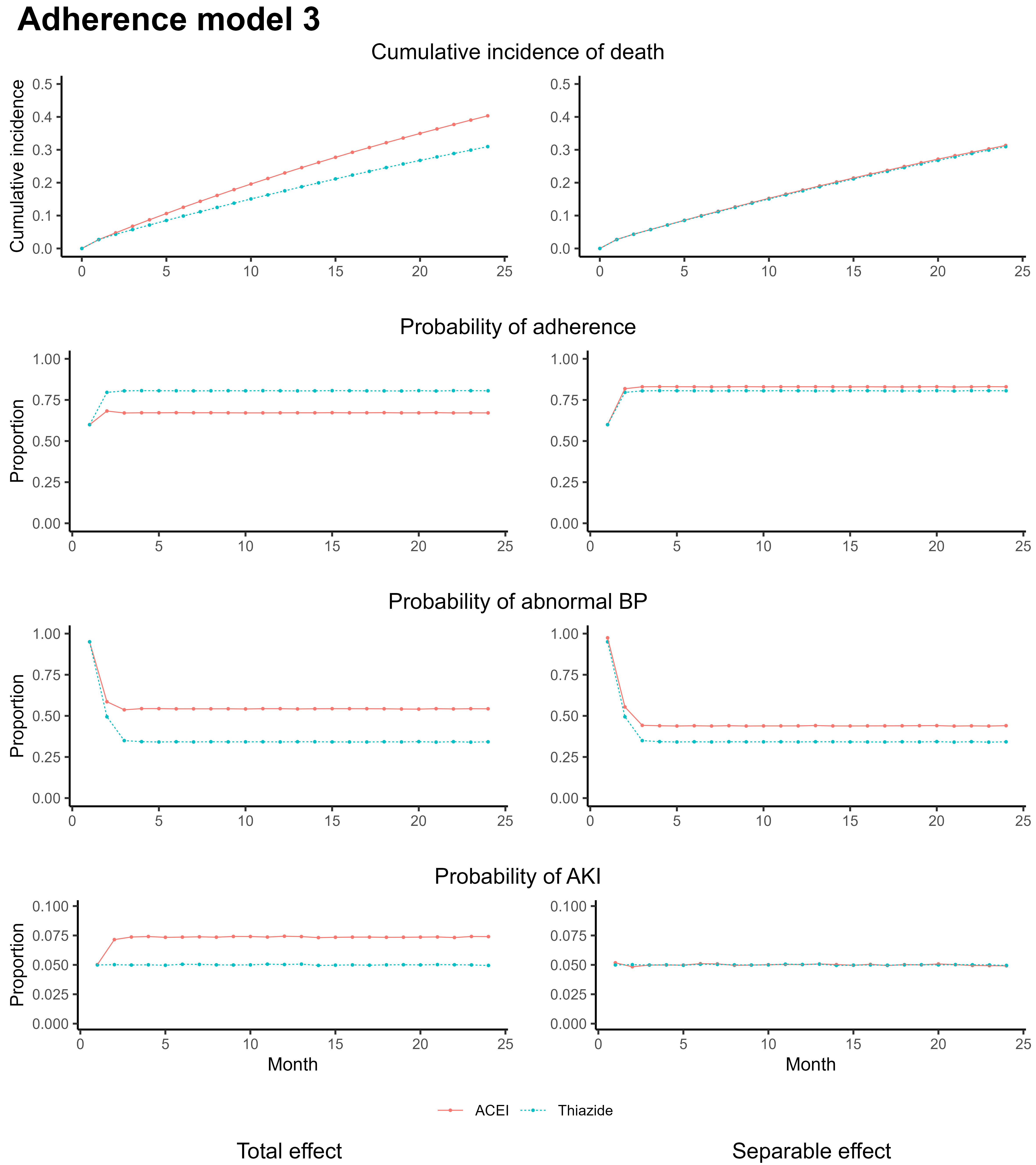}
    \caption{Cumulative incidence of death and probabilities of adherence, AKI, and abnormal blood pressure over 24 months of follow-up for a comparison of ACEI versus thiazide diuretic under the total (`intention-to-treat') effect in the left panels and under the separable effect (comparing a modified version of ACEI to thiazide diuretic) in the right panels. Data were generated using a simulated example with adherence depending only on the $Z_{A}$ component of the initiated medication, past adherence, AKI, and poorly controlled blood pressure.}
    \label{adherence_model_3_results}
\end{figure}

In data generated using the third adherence model, the $Z_{A}$ component of the initiated medication, past adherence, AKI, and poorly controlled blood pressure exert effects on adherence. Under this data generating mechanism, adherence to thiazides is still greater than adherence to ACEIs; but the difference is smaller than in the prior data generating mechanisms because abnormal blood pressure is more likely under ACEI initiation and adherence is higher for those with abnormal blood pressure. Because $Z_{Y}$ exerts an effect on abnormal blood pressure, the separable effect does not balance adherence. Figure~\ref{adherence_model_3_results} shows a small difference in adherence under the separable effect which is a consequence of the effect that the $Z_{Y}$ component has on adherence through its effect on abnormal pressure.

\section{An estimand that prohibits `crossover'}
\label{section: switching prohibited}

\noindent In the main text, we argued that a necessary, but perhaps not sufficient, condition for interpretation of the separable effect of $Z_{Y}$ under $Z_{A}=z_{A}$ as the biological effectiveness of the medications under comparison is that individuals in the study be prohibited from taking a study medication except the one they were assigned at baseline. In some studies, assuming that treatment `crossovers' do not occur might be reasonable. For example, in some randomized trials, individuals do not have access to study medications other than the one they were assigned to at baseline. In observational studies or pragmatic randomized trials where `crossover' is permitted, this assumption will not be plausible. In this section, we describe how the inverse probability weighted representation used to motivate the estimation algorithm in the main text can be extended to identify the separable effect of $Z_{Y}$ under $Z_{A}=z_{A}$ with `crossover' eliminated.

Let $C_{k}$ be an indicator for whether an individual takes a study medication in interval $k$ that they were not assigned to at baseline (e.g. $C_{k} = 1$ if an individual assigned to a thiazide diuretic at baseline takes an angiotensin-converting enzyme inhibitor during interval $k$). An intervention that eliminates `crossover' is analogous to an intervention that eliminates censoring due to loss to follow-up, and the identification strategy for the latter is given by Stensrud et al \citeS{stensrud2021generalizedS}. The results in this section on an estimand that prohibits `crossover' can easily be modified to consider estimands that prohibit censoring due to loss to follow-up or that abolish competing events by re-defining the indicator $C_{k}$ to be an indicator of loss to follow-up or an indicator of the competing event, respectively.

Estimation using inverse probability weighting is motivated by the following weighted representations of the g-formula for the expected counterfactual mean $\mathrm{E}[Y_{K}^{z_{A},z_{Y},\overline{c}_{K}=0}]$:

\begin{align*}
    \sum_{k=1}^{K} \lambda^{C}_{A,k} \prod_{j=1}^{k-1} [1-\lambda^{C}_{A,j}] \text{ or } \sum_{k=1}^{K} \lambda^{C}_{Y,k} \prod_{j=1}^{k-1} [1-\lambda^{C}_{Y,j}]
\end{align*}

where

\begin{align*}
    \lambda^{C}_{A,k} = \frac{\mathrm{E}[Y_k(1-Y_{k-1})W^{C}_{A,k}W^{A}_{C,k} \mid Z=z_{Y}]}{\mathrm{E}[(1-Y_{k-1})W^{C}_{A,k}W^{A}_{C,k} \mid Z=z_{Y}]} \text{ and } \lambda^{C}_{Y,k} = \frac{\mathrm{E}[Y_k(1-Y_{k-1})W^{C}_{Y,k}W^{Y}_{C,k} \mid Z=z_{A}]}{\mathrm{E}[(1-Y_{k-1})W^{C}_{Y,k}W^{Y}_{C,k} \mid Z=z_{A}]} 
\end{align*}

with weights

\begin{align*}
W^{C}_{A,k}= \prod_{j=1}^{k} \Bigg[ &\frac{f(A_{j} \mid \overline{Y}_{j-1}=\overline{C}_{j-1}=0,\overline{L}_{Y,j},\overline{L}_{A,j},\overline{A}_{j-1}, Z=z_{A})}{f(A_{j} \mid \overline{Y}_{j-1}=\overline{C}_{j-1}=0,\overline{L}_{Y,j},\overline{L}_{A,j},\overline{A}_{j-1}, Z=z_{Y})} \times \\
& \frac{\Pr(Z=z_{A} \mid \overline{Y}_{j-1}=\overline{C}_{j-1}=0,\overline{L}_{Y,j-1},\overline{L}_{A,j},\overline{A}_{j-1})}{\Pr(Z=z_{Y} \mid \overline{Y}_{j-1}=\overline{C}_{j-1}=0,\overline{L}_{Y,j-1},\overline{L}_{A,j},\overline{A}_{j-1})} \times \\
& \frac{\Pr(Z=z_{Y} \mid \overline{Y}_{j-1}=\overline{C}_{j-1}=0,\overline{L}_{Y,j-1},\overline{L}_{A,j-1},\overline{A}_{j-1})}{\Pr(Z=z_{A} \mid \overline{Y}_{j-1}=\overline{C}_{j-1}=0,\overline{L}_{Y,j-1},\overline{L}_{A,j-1},\overline{A}_{j-1})}\Bigg],
\end{align*}

\begin{align*}
W^{C}_{Y,k}= & \prod_{j=1}^{k}\frac{f(Y_{j} \mid \overline{Y}_{j-1}=\overline{C}_{j-1}=0, \overline{L}_{Y,j},\overline{L}_{A,j},\overline{A}_{j}, Z=z_{Y})}{f(Y_{j} \mid \overline{Y}_{j-1}=\overline{C}_{j-1}=0, \overline{L}_{Y,j},\overline{L}_{A,j},\overline{A}_{j},Z=z_{A})} \times \\
& \prod_{j=1}^{k} \Bigg[\frac{\Pr(Z=z_{Y} \mid \overline{Y}_{j-1}=\overline{C}_{j-1}=0,\overline{L}_{Y,j},\overline{L}_{A,j},\overline{A}_{j-1})}{\Pr(Z=z_{A} \mid \overline{Y}_{j-1}=\overline{C}_{j-1}=0,\overline{L}_{Y,j},\overline{L}_{A,j},\overline{A}_{j-1})} \times \\
& \hspace{1cm} \frac{\Pr(Z=z_{A} \mid \overline{Y}_{j-1}=\overline{C}_{j-1}=0,\overline{L}_{Y,j-1},\overline{L}_{A,j},\overline{A}_{j-1})}{\Pr(Z=z_{Y} \mid \overline{Y}_{j-1}=\overline{C}_{j-1}=0,\overline{L}_{Y,j-1},\overline{L}_{A,j},\overline{A}_{j-1})}\Bigg],
\end{align*}

\begin{align*}
W^{A}_{C,k}= & \prod_{j=1}^{k}\frac{\mathbbm{1}(C_{j}=0)}{f(C_{j} \mid \overline{Y}_{j-1}=\overline{C}_{j-1}=0, \overline{L}_{Y,j},\overline{L}_{A,j},\overline{A}_{j},Z=z_{Y})},
\end{align*}

and

\begin{align*}
W^{Y}_{C,k}= & \prod_{j=1}^{k}\frac{\mathbbm{1}(C_{j}=0)}{f(C_{j} \mid \overline{Y}_{j-1}=\overline{C}_{j-1}=0, \overline{L}_{Y,j},\overline{L}_{A,j},\overline{A}_{j},Z=z_{A})}.
\end{align*}

Estimates can be obtained using algorithms comparable to those given in the main text and supplement.

%% file: Extra/Figures/adherence_model_1.tex
\begin{figure}
    \centering
\scalebox{0.65}{
\begin{tikzpicture}
\begin{scope}[every node/.style={thick,draw=none}]

    \node (Z)   at ( 0, 0 ) {$Z$};
    \node (ZY)   at ( 1.5, 1 ) {$Z_Y$};
    \node (ZA)   at ( 1.5, -1 ) {$Z_A$};

    \node (LY1)   at ( 3.5, 3 ) {$L_{Y,1}$};
    \node (LY2)  at ( 6.5, 3 ) {$L_{Y,2}$};
    \node (LY3)  at ( 9.5, 3 ) {$L_{Y,3}$};

    \node (LA1)   at ( 3.5, -3 ) {$L_{A,1}$};
    \node (LA2)  at ( 6.5, -3 ) {$L_{A,2}$};
    \node (LA3)  at ( 9.5, -3 ) {$L_{A,3}$};

    \node (A1)   at ( 4, 0 ) {$A_1$};
    \node (A2)  at ( 7, 0 ){$A_2$};
    \node (A3)  at ( 10, 0 ){$A_3$};

    \node (Y1)   at ( 5.5, 0 ) {$Y_{1}$};
    \node (Y2)   at ( 8.5, 0 ) {$Y_{2}$};
    \node (Y3)   at ( 11.5, 0 ) {$Y_{3}$};
    
\end{scope}
\begin{scope}[>={Stealth[black]},
              every node/.style={fill=white,circle},
              every edge/.style={draw=black,very thick}]

    \path [->] (Z)  edge (ZA);
    \path [->] (Z)  edge (ZY);

    \path [->] (ZA)  edge[bend right=15] (A2);
    \path [->] (ZA)  edge[bend right=15] (A3);
    \path [->] (ZA)  edge (LA2);
    \path [->] (ZA)  edge (LA3);

    \path [->] (ZY)  edge (LY2);
    \path [->] (ZY)  edge (LY3);
        
    \path [->] (A1)  edge[bend right=45] (A2);
    \path [->] (A2)  edge[bend right=45] (A3);

    \path [->] (A1)  edge (LA2);
    \path [->] (A2)  edge (LA3);

    \path [->] (A1)  edge (LY2);
    \path [->] (A2)  edge (LY3);

    \path [->] (A1)  edge (Y1);
    \path [->] (A2)  edge (Y2);
    \path [->] (A3)  edge (Y3);

    \path [->] (LA1)  edge (Y1);
    \path [->] (LA2)  edge (Y2);
    \path [->] (LA3)  edge (Y3);

    \path [->] (LY1)  edge (Y1);
    \path [->] (LY2)  edge (Y2);
    \path [->] (LY3)  edge (Y3);
    
\end{scope}
\end{tikzpicture}
}
\caption{Causal DAG representing the causal structure of adherence model 1 at each interval $k={1,2,3}$ for survivors to interval $k$.}
\label{fig: adherence_model_1_dag}
\end{figure}

%% file: Extra/Figures/adherence_model_2.tex
\begin{figure}
    \centering
\scalebox{0.65}{
\begin{tikzpicture}
\begin{scope}[every node/.style={thick,draw=none}]

    \node (Z)   at ( 0, 0 ) {$Z$};
    \node (ZY)   at ( 1.5, 1 ) {$Z_Y$};
    \node (ZA)   at ( 1.5, -1 ) {$Z_A$};

    \node (LY1)   at ( 3.5, 3 ) {$L_{Y,1}$};
    \node (LY2)  at ( 6.5, 3 ) {$L_{Y,2}$};
    \node (LY3)  at ( 9.5, 3 ) {$L_{Y,3}$};

    \node (LA1)   at ( 3.5, -3 ) {$L_{A,1}$};
    \node (LA2)  at ( 6.5, -3 ) {$L_{A,2}$};
    \node (LA3)  at ( 9.5, -3 ) {$L_{A,3}$};

    \node (A1)   at ( 4, 0 ) {$A_1$};
    \node (A2)  at ( 7, 0 ){$A_2$};
    \node (A3)  at ( 10, 0 ){$A_3$};

    \node (Y1)   at ( 5.5, 0 ) {$Y_{1}$};
    \node (Y2)   at ( 8.5, 0 ) {$Y_{2}$};
    \node (Y3)   at ( 11.5, 0 ) {$Y_{3}$};
    
\end{scope}
\begin{scope}[>={Stealth[black]},
              every node/.style={fill=white,circle},
              every edge/.style={draw=black,very thick}]

    \path [->] (Z)  edge (ZA);
    \path [->] (Z)  edge (ZY);

    \path [->] (ZA)  edge[bend right=15] (A2);
    \path [->] (ZA)  edge[bend right=15] (A3);
    \path [->] (ZA)  edge (LA2);
    \path [->] (ZA)  edge (LA3);

    \path [->] (ZY)  edge (LY2);
    \path [->] (ZY)  edge (LY3);
        
    \path [->] (A1)  edge[bend right=45] (A2);
    \path [->] (A2)  edge[bend right=45] (A3);

    \path [->] (A1)  edge (LA2);
    \path [->] (A2)  edge (LA3);

    \path [->] (A1)  edge (LY2);
    \path [->] (A2)  edge (LY3);

    \path [->] (A1)  edge (Y1);
    \path [->] (A2)  edge (Y2);
    \path [->] (A3)  edge (Y3);

    \path [->] (LA1)  edge (A1);
    \path [->] (LA2)  edge (A2);
    \path [->] (LA3)  edge (A3);

    \path [->] (LA1)  edge (Y1);
    \path [->] (LA2)  edge (Y2);
    \path [->] (LA3)  edge (Y3);

    \path [->] (LY1)  edge (Y1);
    \path [->] (LY2)  edge (Y2);
    \path [->] (LY3)  edge (Y3);
    
\end{scope}
\end{tikzpicture}
}
\caption{Causal DAG representing the causal structure of adherence model 2 at each interval $k={1,2,3}$ for survivors to interval $k$.}
\label{fig: adherence_model_2_dag}
\end{figure}

%% file: Extra/Figures/adherence_model_3.tex
\begin{figure}
    \centering
\scalebox{0.65}{
\begin{tikzpicture}
\begin{scope}[every node/.style={thick,draw=none}]

    \node (Z)   at ( 0, 0 ) {$Z$};
    \node (ZY)   at ( 1.5, 1 ) {$Z_Y$};
    \node (ZA)   at ( 1.5, -1 ) {$Z_A$};

    \node (LY1)   at ( 3.5, 3 ) {$L_{Y,1}$};
    \node (LY2)  at ( 6.5, 3 ) {$L_{Y,2}$};
    \node (LY3)  at ( 9.5, 3 ) {$L_{Y,3}$};

    \node (LA1)   at ( 3.5, -3 ) {$L_{A,1}$};
    \node (LA2)  at ( 6.5, -3 ) {$L_{A,2}$};
    \node (LA3)  at ( 9.5, -3 ) {$L_{A,3}$};

    \node (A1)   at ( 4, 0 ) {$A_1$};
    \node (A2)  at ( 7, 0 ){$A_2$};
    \node (A3)  at ( 10, 0 ){$A_3$};

    \node (Y1)   at ( 5.5, 0 ) {$Y_{1}$};
    \node (Y2)   at ( 8.5, 0 ) {$Y_{2}$};
    \node (Y3)   at ( 11.5, 0 ) {$Y_{3}$};
    
\end{scope}
\begin{scope}[>={Stealth[black]},
              every node/.style={fill=white,circle},
              every edge/.style={draw=black,very thick}]

    \path [->] (Z)  edge (ZA);
    \path [->] (Z)  edge (ZY);

    \path [->] (ZA)  edge[bend right=15] (A2);
    \path [->] (ZA)  edge[bend right=15] (A3);
    \path [->] (ZA)  edge (LA2);
    \path [->] (ZA)  edge (LA3);

    \path [->] (ZY)  edge (LY2);
    \path [->] (ZY)  edge (LY3);
        
    \path [->] (A1)  edge[bend right=45] (A2);
    \path [->] (A2)  edge[bend right=45] (A3);

    \path [->] (A1)  edge (LA2);
    \path [->] (A2)  edge (LA3);

    \path [->] (A1)  edge (LY2);
    \path [->] (A2)  edge (LY3);

    \path [->] (A1)  edge (Y1);
    \path [->] (A2)  edge (Y2);
    \path [->] (A3)  edge (Y3);

    \path [->] (LA1)  edge (A1);
    \path [->] (LA2)  edge (A2);
    \path [->] (LA3)  edge (A3);

    \path [->] (LY1)  edge (A1);
    \path [->] (LY2)  edge (A2);
    \path [->] (LY3)  edge (A3);

    \path [->] (LA1)  edge (Y1);
    \path [->] (LA2)  edge (Y2);
    \path [->] (LA3)  edge (Y3);

    \path [->] (LY1)  edge (Y1);
    \path [->] (LY2)  edge (Y2);
    \path [->] (LY3)  edge (Y3);
    
\end{scope}
\end{tikzpicture}
}
\caption{Causal DAG representing the causal structure of adherence model 3 at each interval $k={1,2,3}$ for survivors to interval $k$.}
\label{fig: adherence_model_3_dag}
\end{figure}